\documentclass[a4paper,11pt]{article}
\pdfoutput=1 % if your are submitting a pdflatex (i.e. if you have
\usepackage{jcappub}
\usepackage[T1]{fontenc} % if needed
\usepackage{comment}

\newcommand{\dis}[1]{\begin{equation}\begin{split}#1\end{split}\end{equation}}
\usepackage{xcolor,cancel}

\title{\boldmath Scalar Portal Verifiable Light Dark Matter and Correlated Gravitational Wave Signatures}

\author[a]{Ki-Young Choi,}
\author[a]{Erdenebulgan Lkhagvadorj,}
\author[b]{and Satyabrata Mahapatra}

\affiliation[a]{Department of Physics and Institute of Basic Science, Sungkyunkwan University, 2066 Seobu-ro, Suwon-si, Gyeonggi-do, 16419, Korea}
\affiliation[b]{Indian Institute of Technology Goa,  Ponda-403401, Goa, India}
\emailAdd{kiyoungchoi@skku.edu}
\emailAdd{bulgaa@skku.edu}
\emailAdd{satyabrata@iitgoa.ac.in}

\abstract{ The lack of signals in direct detection experiments has placed the canonical Weakly Interacting Massive Particle (WIMP) paradigm under severe tension, motivating a shift toward the sub-GeV Light Dark Matter (LDM) regime. However, realizing detectable LDM interaction rates typically requires large couplings to the visible sector, which leads to a severe thermal underabundance of the dark matter relic density within standard cosmology. Furthermore, LDM models featuring vector mediators face stringent constraints from the Cosmic Microwave Background (CMB) due to late-time energy injection. In this work, we propose a minimal scalar portal extension featuring a vector-like fermion dark matter candidate, which naturally evades CMB bounds via inherent p-wave annihilation suppression. To simultaneously achieve the correct relic density and large direct-detection couplings, we invoke a pre-Big Bang Nucleosynthesis (BBN) non-standard cosmology dominated by a stiff fluid ($w > 1/3$). The enhanced Hubble expansion during this epoch triggers an early dark matter freeze-out, successfully rescuing the asymptotic relic abundance. Crucially, this stiff pre-BBN phase heavily blue-shifts inflationary gravitational waves that re-enter the horizon prior to BBN, imprinting a distinct high-frequency tilt on the stochastic gravitational wave background. We establish a robust correlation between the non-standard expansion history, the particle physics parameters verifiable in future terrestrial direct detection experiments, and the unique gravitational wave signatures observable by forthcoming space-based interferometers like LISA and DECIGO. This framework highlights how multi-messenger observations can concurrently probe the dark sector and the pre-BBN thermal history of the Universe.
 }

\hypersetup{
colorlinks = true,
linkcolor = blue,
citecolor = magenta
}

\begin{document}
\maketitle
\flushbottom

%%%  ===========================================================================
\section{Introduction}\label{intro}
The pursuit of identifying the fundamental nature of particulate dark matter (DM) stands as one of the foremost challenges in contemporary high-energy physics and cosmology. For several decades, the theoretical and experimental landscape has been overwhelmingly dominated by the Weakly Interacting Massive Particle (WIMP) paradigm~\cite{Kolb:1990vq}. Its enduring appeal stems from the so-called "WIMP miracle", whereby a stable particle with an electroweak-scale mass and interaction strength naturally freezes out of the primordial thermal bath, yielding a relic density ($\Omega_{\text{DM}} h^2 \simeq 0.12$~\cite{Planck:2018vyg,ParticleDataGroup:2020ssz}) in remarkable agreement with cosmological observations~\cite{Steigman:2012nb}. Despite its theoretical success, however, the WIMP paradigm is currently experiencing an unprecedented crisis~\cite{Arcadi:2017kky, Roszkowski:2017nbc, Arcadi:2024ukq}. The persistent absence of definitive signals across a global network of highly sensitive terrestrial direct detection experiments, particularly large-scale liquid noble gas time projection chambers such as LZ~\cite{LZ:2024zvo}, XENONnT~\cite{XENON:2024wpa}, and PandaX-4T~\cite{PandaX:2025rrz}, has placed the canonical electroweak-scale WIMP under severe tension. As the allowable parameter space for traditional WIMPs is increasingly squeezed toward the coherent elastic neutrino-nucleus scattering boundary commonly referred to as the neutrino floor~\cite{Hertel:2018aal}, the focus of the field has necessarily expanded beyond the traditional WIMP framework. 

Growing theoretical interest and rapidly advancing experimental capabilities have established the sub-GeV Light Dark Matter (LDM) regime~\cite{COSINE-100:2021poy,XENON:2019gfn,CRESST:2019jnq,XENON:2024znc,PandaX-II:2021nsg,PandaX:2022xqx,SENSEI:2020dpa,SuperCDMS:2024yiv,DAMIC-M:2025luv} as a highly compelling and phenomenologically rich alternative landscape~\cite{Balan:2024cmq,Cheek:2025nul,Krnjaic:2025noj,Dutta:2019fxn,Essig:2017kqs,Bondarenko:2019vrb,Adhikary:2024btd,Borah:2024yow,Borah:2025wcc}.
However, transitioning to lower mass scales introduces unique theoretical and observational challenges. In particular, sub-GeV DM particles transfer exceedingly small kinetic energy during elastic scattering with target nuclei. Because the momentum transfer scales with the reduced mass of the system, LDM scattering fails to exceed the stringent recoil energy thresholds of conventional nuclear recoil detectors~\cite{GlobalArgonDarkMatter:2022ppc,Schumann:2019eaa}. Consequently, standard direct detection techniques become largely insensitive in this mass range. To maintain the possibility of direct experimental verification, LDM models strictly require the presence of a relatively light mediator particle that couples the dark sector to the Standard Model (SM) visible sector as achieving observable event rates in current and next-generation direct detection experiments necessitates large effective couplings between the dark matter and the visible sector~\cite{Elor:2021swj, Zurek:2024qfm}.

This phenomenological requirement immediately precipitates two severe cosmological tensions. First, substantial couplings combined with a light mediator mass induce an exceptionally efficient annihilation cross-section for the dark matter particles in the early Universe. Within the standard radiation-dominated (RD) cosmological history, this highly efficient annihilation mechanism maintains the dark matter in thermal equilibrium for an extended duration. As a result, the DM abundance experiences prolonged exponential Boltzmann suppression, culminating in a severe underabundance of the final relic density~\cite{Elor:2021swj, Baer:2014eja}. Second, such scenario faces stringent constraints from the Cosmic Microwave Background (CMB)~\cite{Slatyer:2015jla, Elor:2015bho} and indirect detection experiments~\cite{Fermi-LAT:2015att, HESS:2018cbt, Profumo:2017obk}. Dark matter annihilating during the recombination epoch ($z \sim 1100$) injects high-energy SM particles into the intergalactic medium, significantly altering the ionization history of the Universe. For s-wave annihilating thermal relics, precise CMB measurements from the Planck satellite firmly rule out dark matter masses below approximately 20 GeV~\cite{Madhavacheril:2013cna, Slatyer:2015jla, Elor:2015bho}.

Motivated by the necessity of possessing an enhanced direct detection rate while yielding the correct relic density and simultaneously avoiding fatal CMB constraints, this report investigates a non-standard cosmology (NSC) framework coupled with a scalar-mediated dark sector scenario. To bypass the CMB limits, the dark sector is modeled with a vector-like singlet fermion as the DM candidate and a light scalar mediator. Due to strict angular momentum selection rules, the annihilation of this fermion into a pair of scalar mediators is entirely $p$-wave suppressed. Because the relative velocity of dark matter during recombination is infinitesimally small, the late-time annihilation rate vanishes, cleanly evading the CMB bounds.

To resolve the relic density underabundance, a principled departure from standard cosmological assumptions is required. Non-standard cosmologies are highly viable, as the expansion history of the Universe prior to Big Bang Nucleosynthesis (BBN) remains entirely unconstrained by direct observation~\cite{Redmond:2017tja, DEramo:2017gpl, Visinelli:2017qga}. If the Universe underwent a pre-BBN epoch dominated by a fluid with a "stiff" equation of state (EoS, $w > 1/3$), such as a kination era, the Hubble expansion rate would be drastically enhanced~\cite{Arcadi:2024jzv, Arias:2019uol}. This higher temperature enhanced expansion forces the dark matter to freeze out prematurely, effectively arresting the efficient annihilation process and improving the final abundance of the relic to match the observed values.

Crucially, the invocation of a stiff non-standard cosmological epoch is not merely an ad-hoc theoretical consideration; rather, it inherently leaves a measurable, tightly correlated signature in the primordial stochastic gravitational wave (GW) background~\cite{Soman:2024zor, Mishra:2025nnu, Konings:2024zvz, Ghoshal:2025ldb, Kuroyanagi:2014nba}. Inflationary gravitational waves, which are typically scale-invariant, re-enter the horizon and experience a modified transfer function depending on the dominant energy fluid. Modes re-entering during a stiff era are heavily blue-shifted, imprinting a distinct spectral tilt that serves as a direct measure of the NSC's equation of state.  
Such a signal may be accessible to future GW observatories such as aLIGO~\cite{LIGOScientific:2014pky}, Einstein Telescope (ET)~\cite{ET:2019dnz}, Cosmic Explorer (CE)~\cite{Reitze:2019iox}, $\mu$-ARES~\cite{Sesana:2019vho} and THEIA~\cite{Garcia-Bellido:2021zgu}, thereby providing an indirect probe of the pre-BBN expansion history. Consequently, the cosmological parameters governing the stiff epoch, most notably the equation-of-state parameter $w$ and the reheating temperature $T_{\rm rh}$, simultaneously determine the dark matter relic abundance and the spectral features of the stochastic GW background. 
 {Thus, a comprehensive GW tomography of the NSC not only validates the cosmological history required to rescue LDM but also establishes a rigid correlation between the viable particle physics parameters (DM mass, mediator mass, dark sector coupling), the cosmological parameters (the EoS parameter and the reheating temperature), and the Signal-to-Noise Ratio (SNR) in future GW observatories.}
This establishes a direct connection between dark matter phenomenology and gravitational-wave observations, offering a complementary avenue to test the underlying non-standard cosmological history.

The paper is organized as follows. In section~\ref{sec:dmpheno}, we review basics of the LDM model, discuss its prospect for direct detection, and identify the parameter space consistent with the current experimental constraints while remaining accessible to future searches. We also describe the dark matter relic abundance in the context of the non-standard cosmological history. In section~\ref{sec:gw}, we review the inflationary GW background and provide both analytic and numerical estimation for its enhancement during the fluid-$\xi$ dominated epoch. Finally, we summarize our results and conclude in section~\ref{sec:conc}.

\section{Dark Matter Phenomenology }\label{sec:dmpheno}

\subsection{A Minimal Framework}
We propose a minimal, renormalizable extension to the SM, introducing a highly constrained dark sector with a vector-like singlet fermion $\psi$, which serves as the primary dark matter candidate, and a real singlet scalar field $\Phi$, which mediates interactions between the dark and the visible sectors. The choice of a vector-like fermion allows for a gauge-invariant bare Dirac mass term without relying on spontaneous symmetry breaking in the dark sector, while the singlet nature of the fields ensures that the model inherently introduces no new SM gauge anomalies. The relevant terms in the Lagrangian characterizing the dark sector and its portal to the visible SM sector are given by:
\begin{align}
    \mathcal{L} \supset i \bar{\psi} \gamma^\mu \partial_\mu \psi - m_\psi \bar{\psi} \psi - y_\psi \bar{\psi} \Phi \psi +{\it h.c.} - V(\Phi, H) \,,
\end{align}
where $m_\psi$ represents the bare mass of the vector-like fermion, and $y_\psi$ designates the Yukawa coupling between the dark matter and the singlet scalar mediator $\Phi$. The vector-like nature of $\psi$ permits the bare mass term $m_\psi \bar{\psi} \psi$ prior to any spontaneous symmetry breaking. Furthermore, the stability of the dark matter particle is ensured by an imposed $\mathbb{Z}_2$ symmetry under which $\psi \to -\psi$, whereas the scalar mediator $\Phi$ and the SM Higgs doublet $H$ are even.
The scalar potential $V(\Phi, H)$ dictates the vacuum structure, the mass generation for the scalar sector, and the critical mixing between the dark mediator and the SM Higgs doublet $H$. The most general renormalizable scalar potential compatible with the gauge symmetries of the SM can be written as:
\begin{align}
       V(H, \Phi)&= -\mu^2_H \left(H^\dagger H \right) + \lambda_H \left(H^\dagger H \right)^2 -\mu^2_{\Phi} \left(\Phi^\dagger \Phi\right) + \lambda_\Phi \left(\Phi^\dagger \Phi \right)^2  \nonumber\\&+ \lambda_{H \Phi} \left(H^\dagger H\right)\left(\Phi^\dagger \Phi\right)-\mu_{ H\Phi} \Phi  \left(H^\dagger H\right)\,+\frac{1}{3}\mu_3 \Phi^3+{\it h.c.}
\end{align}

For the purpose of minimal dark matter phenomenology, the parameters that dictate the portal interactions are the dimensionless quartic coupling $\lambda_{H \Phi}$ and the dimensionful trilinear coupling $\mu_{H \Phi}$.  {While a general renormalizable potential allows for the cubic self-coupling $\frac{1}{3}\mu_3 \Phi^3$, we assume for simplicity that this parameter is negligibly small ($\mu_3 \simeq 0$). This choice minimizes the number of free parameters without altering the primary dark matter interactions of interest.}\\

Upon electroweak symmetry breaking, the neutral component of the Higgs doublet acquires a vacuum expectation value (VEV), denoted as $v_H \simeq 246$ GeV. Concurrently, the singlet scalar $\Phi$ may also acquire a non-zero VEV, $v_\Phi$. We expand the scalar fields around their respective VEVs:
\begin{equation}H = \begin{pmatrix} 0 \ \frac{v_H + h}{\sqrt{2}} \end{pmatrix}, \quad \Phi = v_\Phi + \phi \,.\end{equation}
Substituting these expansions back into the scalar potential reveals that the interaction eigenstates $h$ and $\phi$ are not the physical mass eigenstates. The trilinear term $\mu_{H \Phi}$ and the portal coupling $\lambda_{H \Phi}$ (in conjunction with the VEVs) induce an off-diagonal mass mixing matrix for the CP-even scalars. The physical mass eigenstates, denoted conventionally as $h_1$ (which we identify as the SM-like Higgs boson with $m_{h_1} \simeq 125$ GeV) and $h_2$ (the new light scalar mediator), are obtained via an orthogonal rotation parameterized by a mixing angle $\theta$:
\begin{equation}\begin{pmatrix} h_1 \\ h_2 \end{pmatrix} = \begin{pmatrix} \cos\theta & \sin\theta \\ -\sin\theta & \cos\theta \end{pmatrix} \begin{pmatrix} h \\ \phi \end{pmatrix}\,.
\end{equation} 
Therefore, the corresponding mass matrix is found as: 
\dis{
M^2 = \begin{pmatrix} 2 \lambda_H \upsilon_H^2 & \upsilon_H (2 \lambda_{H \Phi} \upsilon_\Phi - \mu_{H \Phi})  \\ \upsilon_H (2 \lambda_{H \Phi} \upsilon_\Phi - \mu_{H \Phi}) & 8 \lambda_\Phi \upsilon_\Phi^2 \end{pmatrix}\,.
}
The mixing angle $\theta$ is explicitly determined by the fundamental parameters of the scalar potential:
\begin{equation}
\tan(2\theta) = \frac{ 2 \lambda_{H \Phi} v_H v_\Phi - \mu_{H \Phi} v_H}{\lambda_{H} v_H^2 - 4 \lambda_{\Phi} v_\Phi^2 } =   \frac{ 4 \lambda_{H \Phi} v_H v_\Phi - 2 \mu_{H \Phi} v_H}{m_h^2 - 8m_\phi^2}\,.
\end{equation} 
For a sub-GeV scalar mediator, this mixing angle $\theta$ is severely constrained by a multitude of experiments as well as cosmological and astrophysical observations.  
Nevertheless, a nonzero mixing angle $\theta$ induces couplings of the light scalar mediator $h_2$ to nucleons through its Higgs admixture, thereby enabling interactions between the dark matter particle $\psi$ and ordinary matter. As a result, $\psi$ can scatter elastically off nuclei via the $t$-channel exchange of the scalar mediators, providing the primary experimental signature of the model in direct detection searches.
The dark matter particle $\psi$ interacts with a nucleon $N$ ($N = p, n$) via the $t$-channel exchange of the mixed scalars $h_1$ and $h_2$. The spin-independent (SI) dark matter-nucleon elastic scattering cross-section is expressed analytically as:
\begin{equation}\sigma_{\psi N}= \frac{\mu^2_{\psi N}}{\pi} \frac{\left(y_\psi~ y^{\rm eff}_{\phi N}\right)^2}{\left[m^2_{h_2} + m^2_\psi v_\psi^2\right]^2}\,,
\end{equation}
where $\mu_{\psi N} = m_\psi m_N / (m_\psi + m_N)$ is the reduced mass of the dark matter-nucleon system and we take $m_\psi > m_{h_2}$. The term $m_\psi^2 v_\psi^2$ characterizes the typical scale of the squared momentum transfer $q^2$ (since $q \sim \mu_{\psi N} v_\psi \approx m_\psi v_\psi$ in the light dark matter limit). For the typical galactic escape velocities governing the local dark matter halo ($v_\psi \sim 10^{-3} c$), the momentum transfer is small. If we assume a mediator mass $m_{h_2}$ that is large compared to the momentum transfer ($m_{h_2} \gg q$), the propagator denominator simplifies to $m_{h_2}^4$.
The effective coupling to the nucleons, $y^{\rm eff}_{\phi N}$, is derived from the scalar mediator's mixing with the Higgs and the Higgs-nucleon interaction strength:\begin{equation}y^{\rm eff}_{\phi N} = \sin\theta  \frac{m_N}{v_H} f_N \simeq 1.2 \times 10^{-3} \sin\theta\,,
\end{equation}
where $f_N = \sum_{q=u,d,s} f_{Tq}^{(N)} + \frac{2}{9} f_{TG}^{(N)} \simeq 0.3$ is the effective nuclear form factor parameterizing the Higgs coupling to the nucleon, derived from lattice QCD and chiral perturbation theory~\cite{Hoferichter:2017olk, Bertone:2004pz, Ellis:2000ds}.

\begin{figure}[tbp]
\centering 
\includegraphics[width=.7\textwidth]{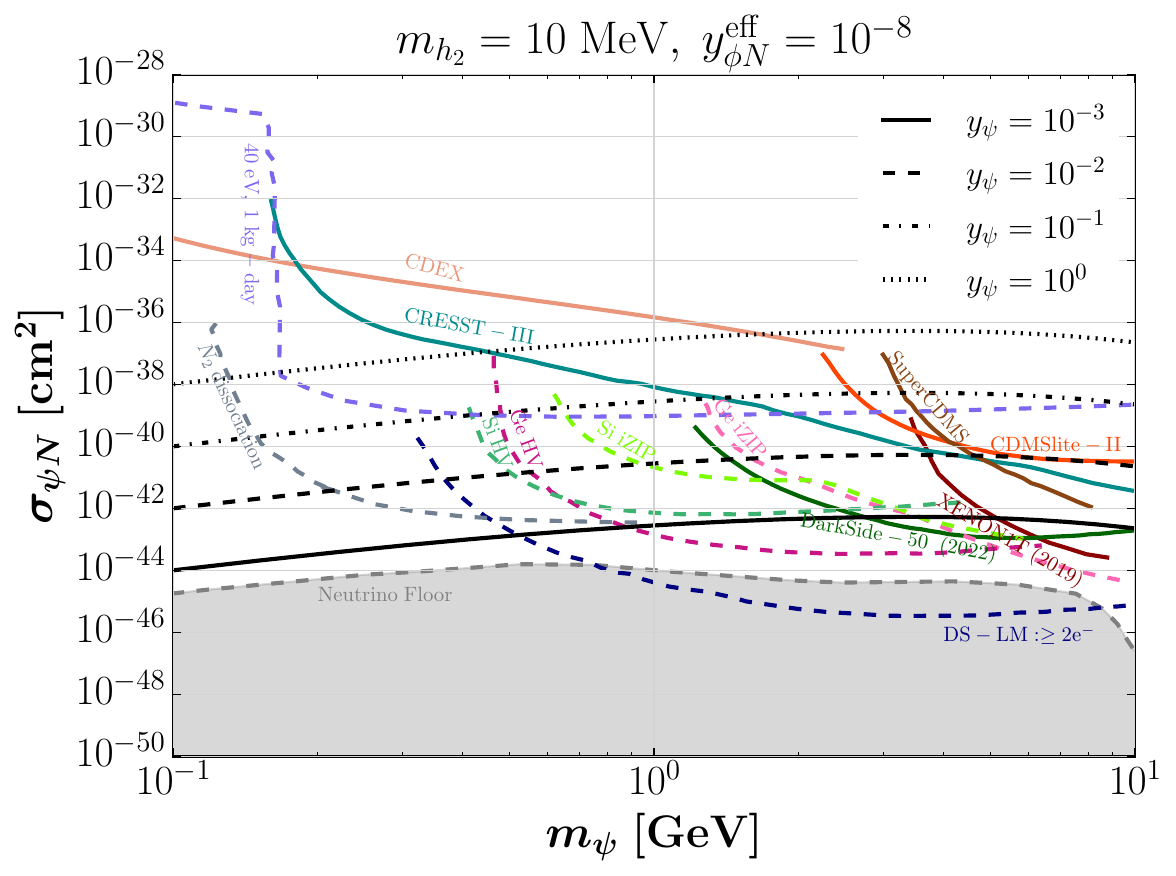}
\caption{Predicted spin-independent dark matter-nucleon scattering cross section, $\sigma_{\psi N}$, as a function of the dark matter mass $m_\psi$ for representative values of the Yukawa coupling $y_\psi$, shown by the black curves. The results are compared with current 90\% C.L. upper limits (solid lines) and the projected sensitivities of future direct detection experiments (dashed lines), explicitly listed in the main text. We set the scalar mediator mass to $m_{h_2} = 10 \ \rm{MeV}$ and the effective scalar-nucleon coupling to $y_{\phi N}^{\rm eff} = 10^{-8}$. }
\label{fig:scattering_crossSection}
\end{figure}

In Fig.~\ref{fig:scattering_crossSection}, we present the $\sigma_{\psi N}$ estimate for DM masses in the range $100 \ \rm{MeV} - 10 \ \rm{GeV}$ and several representative values of the Yukawa coupling $y_\psi$. Throughout this figure, we fix the scalar mediator mass and the effective scalar-nucleon coupling to $m_{h_2} = 10 \ \rm{MeV}$ and $y_{\phi N}^{\rm eff} = 10^{-8}$, respectively. The current strongest constraints from the XENON1T collaboration~\cite{XENON:2018voc, XENON:2019gfn}, DarkSide-50~\cite{DarkSide-50:2022qzh}, SuperCDMS~\cite{SuperCDMS:2015eex},  CDMSlite~\cite{SuperCDMS:2013eoh}, CRESST-III~\cite{CRESST:2019jnq}, and CDEX~\cite{CDEX:2019hzn} are shown as solid curves. The gray shaded region, labeled as neutrino floor, indicates the WIMP-discovery limit from~\cite{Hertel:2018aal}, extended to lower masses for $^4\rm{He}$-based experiments. In addition, the projected HeRALD sensitivity to the DM-nucleon SI interaction is shown by dashed slate-blue, assuming an exposure of 1 kg-day and an energy threshold of 40 eV~\cite{Hertel:2018aal}. Furthermore, we include the projected sensitivities of SuperCDMS SNOLAB experiment~\cite{SuperCDMS:2016wui}, which targets sub-$10 \ \rm{GeV}$ dark matter using cryogenic HV and iZIP detectors with germanium and silicon target materials. Finally, we display the projected sensitivities corresponding to three signal events from the dissociation of $N_2$ molecules with an exposure of 1 kg-yr~\cite{Alexander:2016aln} (gray  dashed line) and from an experiment employing liquid argon with $^{39}\rm{Ar}$ activity of $73 \ \mu {\rm{Bq}/kg}$ and $2 e^-$ threshold~\cite{GlobalArgonDarkMatter:2022ppc} (navy dashed line). 

\begin{figure}[tbp]
\centering 
\includegraphics[width=.49\textwidth]{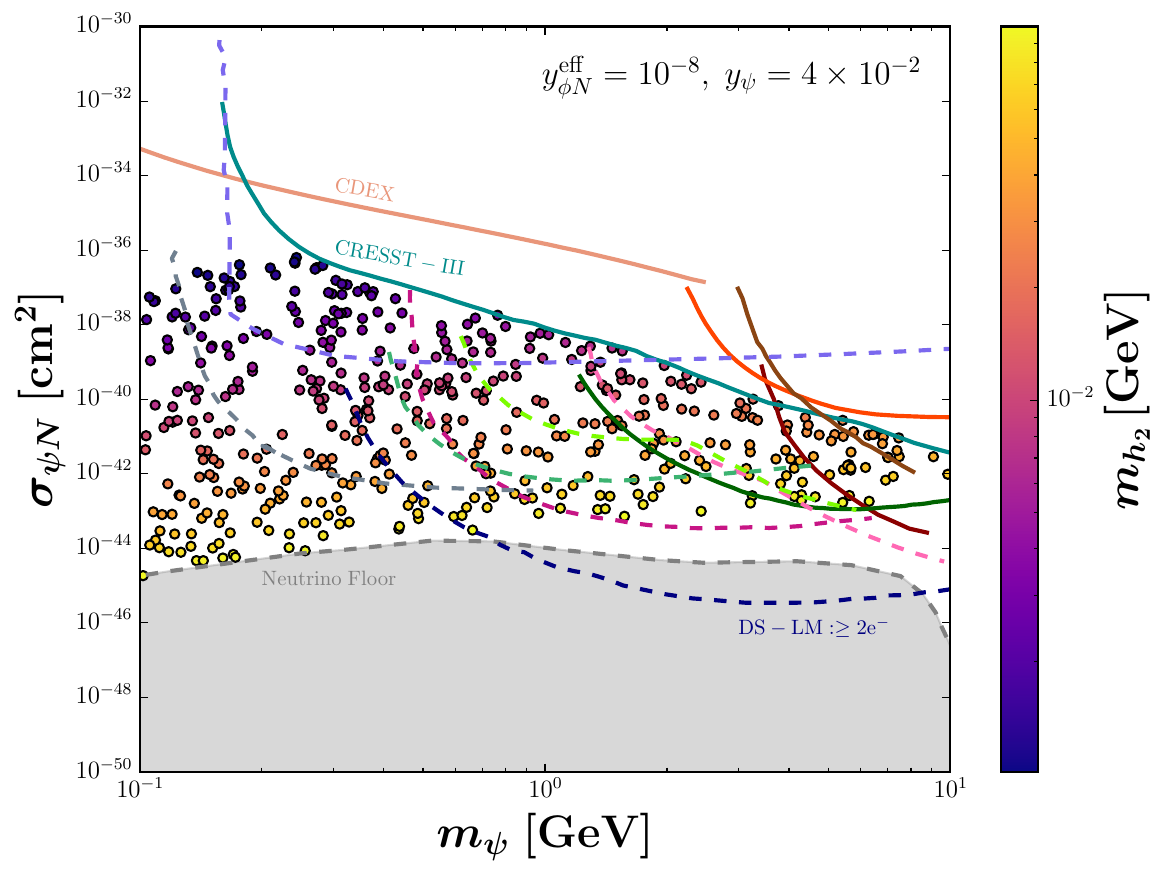}
\hfill
\includegraphics[width=.49\textwidth]{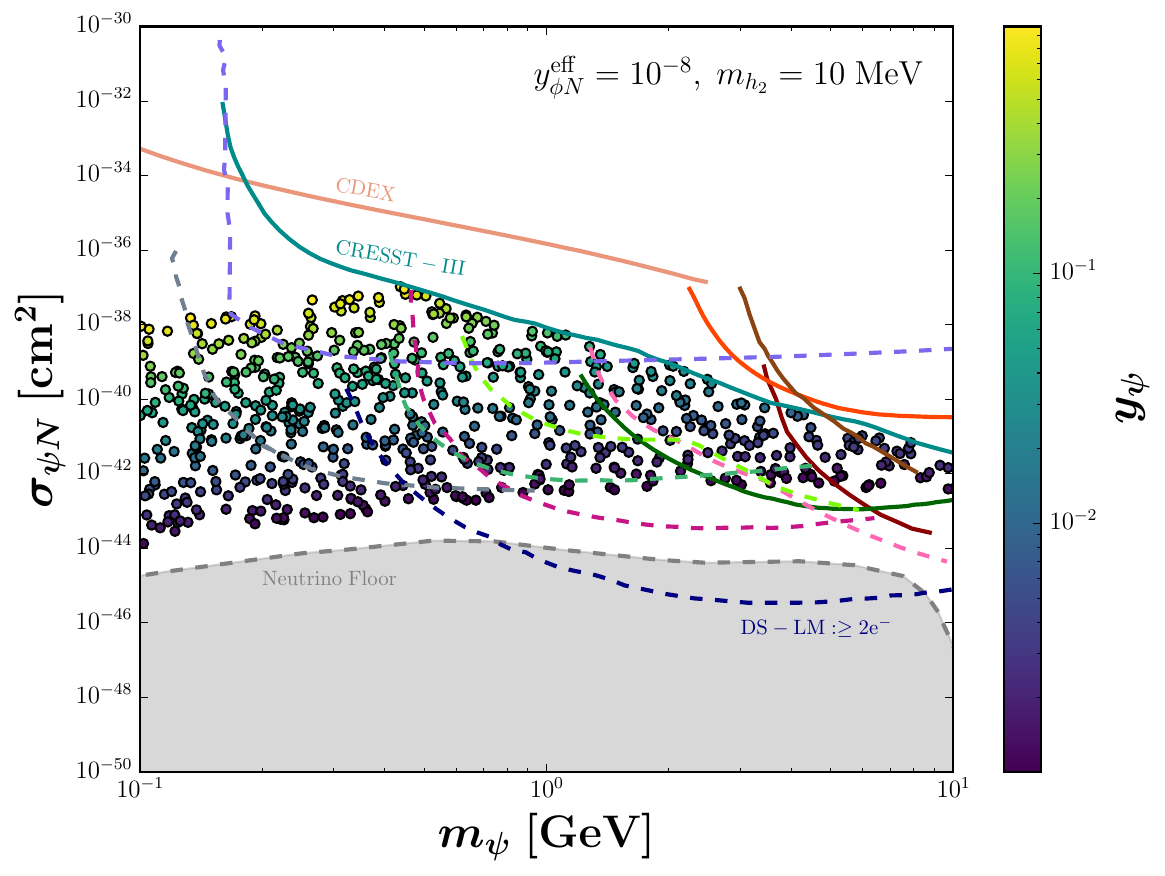}
\caption{Parameter space in the plane of the spin-independent DM-nucleon scattering cross section and the DM mass $m_\psi$. The displayed points satisfy the current CRESST-III direct detection bound, with the color coding representing the mediator mass $m_{h_2}$ for $y_\psi = 4 \times 10^{-2}$ (left) and the Yukawa coupling $y_\psi$ for $m_{h_2} = 10 \ \rm{MeV}$ (right). Throughout, we set $y_{\phi N}^{\rm eff} = 10^{-8}$.}
\label{fig:crossSection_paramScan}
\end{figure}

 In Fig.~\ref{fig:crossSection_paramScan}, we showcase the spin-independent DM-nucleon scattering cross section, $\sigma_{\psi N}$ as a function of the DM mass $m_\psi$, scrutinizing against the most stringent current DM direct detection constraint from CRESST-III~\cite{CRESST:2019jnq}. In order to determine the parameter region compatible with DM direct detection upper limit, we conduct a random scan over the ranges $m_\psi \in [100 \ \rm{MeV}, 10 \ \rm{GeV}]$, $m_{h_2} \in [1 \ \rm{MeV}, 100 \ \rm{MeV}]$, and $y_{\psi} \ \in [10^{-3},1]$, while fixing the effective mediator-nucleon coupling to $y_{\phi N}^{\rm eff} = 10^{-8}$. 
 For visualization purpose, we display only two representative subsets of the scan. In the left panel, the Yukawa coupling is fixed to $y_\psi = 4 \times 10^{-2}$, and use the color scale indicates the corresponding mediator mass, $m_{h_2}$. In the right panel, we fix $m_{h_2}=10~\mathrm{MeV}$, with the color scale representing the Yukawa coupling, $y_\psi$. In both panels, only parameter points that lie below the current CRESST-III exclusion limit are shown. The remaining experimental constraints and projected sensitivities are identical to those presented in Fig.~\ref{fig:scattering_crossSection}.

To maximize the prospects for future experimental detection while remaining consistent with current direct detection constraints, the effective dark sector coupling, $(y_\psi \sin\theta)$, must lie close to its current upper bound. However, such a large coupling, together with a light scalar mediator, leads to highly efficient dark matter annihilation into mediator pairs, yielding a relic abundance well below the observed value in the standard RD cosmology. This tension motivates the consideration of the NSC, in which dark matter freezes out during an epoch with an enhanced Hubble expansion rate. The faster expansion causes freeze-out to occur earlier, reducing the annihilation efficiency and restoring the observed relic abundance. Remarkably, such non-standard thermal histories can be probed through the stochastic gravitational wave background. We discuss the resulting cosmological evolution and its gravitational-wave signatures in the following sections.

\subsection{Dark Matter Relic Density in Non-Standard Cosmology}
In the early Universe, the vector-like dark matter fermion $\psi$ maintains thermal equilibrium with the SM bath through the scalar portal interactions. The dark matter can annihilate into SM particles via $s$-channel exchange of the mixed scalars $h_1$ and $h_2$. However, in the hierarchical parameter space relevant for light dark matter, where $m_\psi > m_{h_2}$, the dark sector annihilation channel $\bar{\psi}\psi \to h_2 h_2$ via $t$-channel and $u$-channel $\psi$ exchange dominates the total annihilation cross-section. 

A defining phenomenological feature of this specific model is the velocity dependence of the annihilation cross-section into the mediator pair. To understand this, we expand the thermally averaged annihilation cross-section into partial waves:
$\langle \sigma v \rangle = a + b \langle v^2 \rangle + \mathcal{O}(v^4)$
where the coefficient $a$ represents the $s$-wave contribution (velocity independent), and the coefficient $b$ represents the $p$-wave contribution.

For a Dirac or vector-like fermion annihilating into a pair of real scalars, the initial state $\bar{\psi}\psi$ must conform to strict selection rules governed by parity ($P$) and charge conjugation ($C$). Assuming CP is conserved in the dark sector, the $s$-wave amplitude ($L=0$) strictly vanishes. This is because an $s$-wave fermion-antifermion pair has $P = (-1)^{L+1} = -1$, whereas a pair of identical real scalars inherently possesses $P = +1$. To match quantum numbers, the initial state must carry orbital angular momentum $L=1$, corresponding to a $p$-wave process. Consequently, the leading-order annihilation cross-section is entirely governed by the $p$-wave term and can be expressed analytically as:

\begin{equation} \label{eq:crossSec_annihilation}
\langle \sigma v \rangle_{\bar{\psi}\psi \to h_2 h_2} \simeq \frac{3 y_\psi^4}{64 \pi m_\psi^2} \frac{(1 - m_{h_2}^2/m_\psi^2)^{3/2}}{(1 - m_{h_2}^2/2m_\psi^2)^4} v^2,
\end{equation}
where $v$ is the relative velocity of the annihilating dark matter particles. Within the thermal bath, the thermally averaged velocity squared is related to the temperature by $\langle v^2 \rangle \simeq 6T/m_\psi$. Thus, the cross-section scales linearly with temperature in the non-relativistic limit: $\langle \sigma v \rangle \simeq \sigma_1 (T/m_\psi)$.

The magnitude of the Yukawa coupling $y_\psi$ required to satisfy direct detection observability creates an unacceptably large cross-section at the time of freeze-out. 
In a standard cosmological history, dark matter decouples when the expansion rate of the Universe, $H$, exceeds the annihilation rate, $\Gamma = n_\psi \langle \sigma v \rangle$. If $\langle \sigma v \rangle$ is extraordinarily large, the freeze-out is delayed to much lower temperatures. Because the equilibrium number density drops exponentially as $n_{\psi,\rm{eq}} \propto e^{-m_\psi/T}$ for non-relativistic particles, delaying freeze-out means the exponential suppression operates for a longer duration. Consequently, the resulting relic density $\Omega_{\text{DM}} h^2$ falls to values far below the correct relic density, rendering the model underabundant.

To resolve this underabundance, the expansion history of the early Universe must be altered. The standard cosmological model posits that the energy density between the end of post-inflationary reheating and BBN is entirely dominated by a radiation fluid, possessing an equation of state $w = p/\rho = 1/3$. If, instead, the Universe underwent the NSC phase dominated by an alternate fluid or scalar field with an equation of state satisfying $w > 1/3$, the thermodynamics change profoundly. The upper theoretical limit is $w=1$, corresponding to an era where the kinetic energy of a fast-rolling field completely eclipses its potential energy, {an epoch called "kination".} During such a non-standard cosmological era, the total energy density of the Universe is governed by $\rho_{\text{tot}} = \rho_R + \rho_\xi$, where $\rho_R$ is the subdominant radiation energy density and $\rho_\xi$ is the dominant energy density. The evolution of the dark matter number density $n_\psi$ alongside the dynamic, transitioning cosmological background requires solving a system of coupled Boltzmann equations. We assume that the fluid-$\xi$ does not decay nor interact with dark matter or radiation. On the other hand, radiation and dark matter are thermally coupled via pair production and annihilation. Therefore, the differential equations governing the system are~\cite{Redmond:2017tja}:

\begin{align}
&\frac{d\rho_\xi}{dt} + 3H(1+w)\rho_\xi = 0\,, \label{eq:rhoxi}\\
&\frac{d\rho_R}{dt} + 4H\rho_R = 2 \langle \sigma v \rangle \langle E_\psi \rangle \left( n_\psi^2 - n_{\psi,\text{eq}}^2 \right) \,, \\
%\Gamma_\xi \rho_\xi
&\frac{dn_\psi}{dt} + 3Hn_\psi = - \langle \sigma v \rangle \left( n_\psi^2 - n_{\psi,\text{eq}}^2 \right) \,,
\end{align}
where  $\langle E_\psi \rangle $ is the average energy of a dark matter particle and $ n_{\psi,\text{eq}}$ is the number density of dark matter particles in thermal equilibrium, $n_{\psi,\text{eq}} = \frac{g_\psi}{2 \pi^2} m_\psi^2 T K_2(m_\psi/T)$, $T$ being the temperature of the Standard Model radiation bath.
Solving Eq.~(\ref{eq:rhoxi}), one can find the straightforward solution for the fluid-$\xi$ energy density as
\dis{
\rho_\xi (T) \simeq \rho_{\xi} (T_i) \left(\frac{a_i}{a}\right)^{3 (1+w)} = \rho_{\xi} (T_i) \left(\frac{g_{*,s}(T)}{g_{*,s}(T_{i})} \right)^{(1+w)} \left(\frac{T}{T_i}\right)^{3(1+w)}\,,
\label{rho_xi_T}
}
where we have assumed that entropy conservation ensures $g_{*,s}(T)^{1/3} T a= \rm{const}$ in the last step. The contribution from radiation can be expressed in terms of its temperature as follows 
\dis{
\rho_R(T) = \frac{\pi^2}{30} g_*(T) T^4\,.
}

 {At the reheating temperature $T_{\rm rh}$, we require the energy density of the fluid-$\xi$ to equal that of radiation, namely $\rho_\xi(T_{\rm rh}) = \rho_R(T_{\rm rh})$. Using Eq.~(\ref{rho_xi_T}), this condition fixes the initial energy density $\rho_\xi(T_i)$ in terms of the reheating temperature $T_{\rm rh}$, the initial temperature $T_i$, and the EoS parameter $w$:
\dis{\label{eq:rho_xi_Tin}
\rho_\xi(T_i) = \frac{\pi^2}{30} g_*(T_{\rm rh}) T_{\rm rh}^4 \left(\frac{g_{*,s}(T_{i})}{g_{*,s}(T_{\rm rh})} \right)^{1+w} \left(\frac{T_i}{T_{\rm rh}}\right)^{3(1+w)}\,.}}

\begin{figure}[tbp]
\centering 
\includegraphics[width=.49\textwidth]{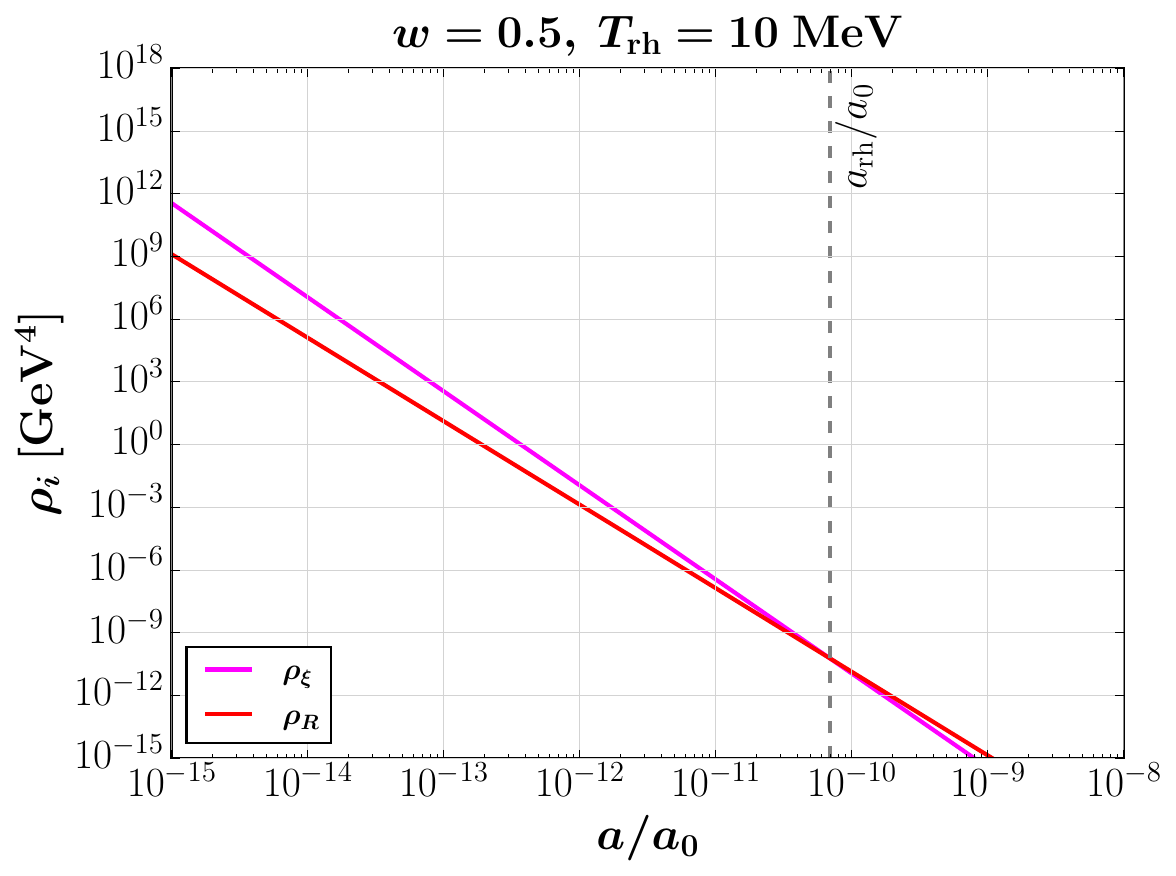}
\hfill
\includegraphics[width=.49\textwidth]{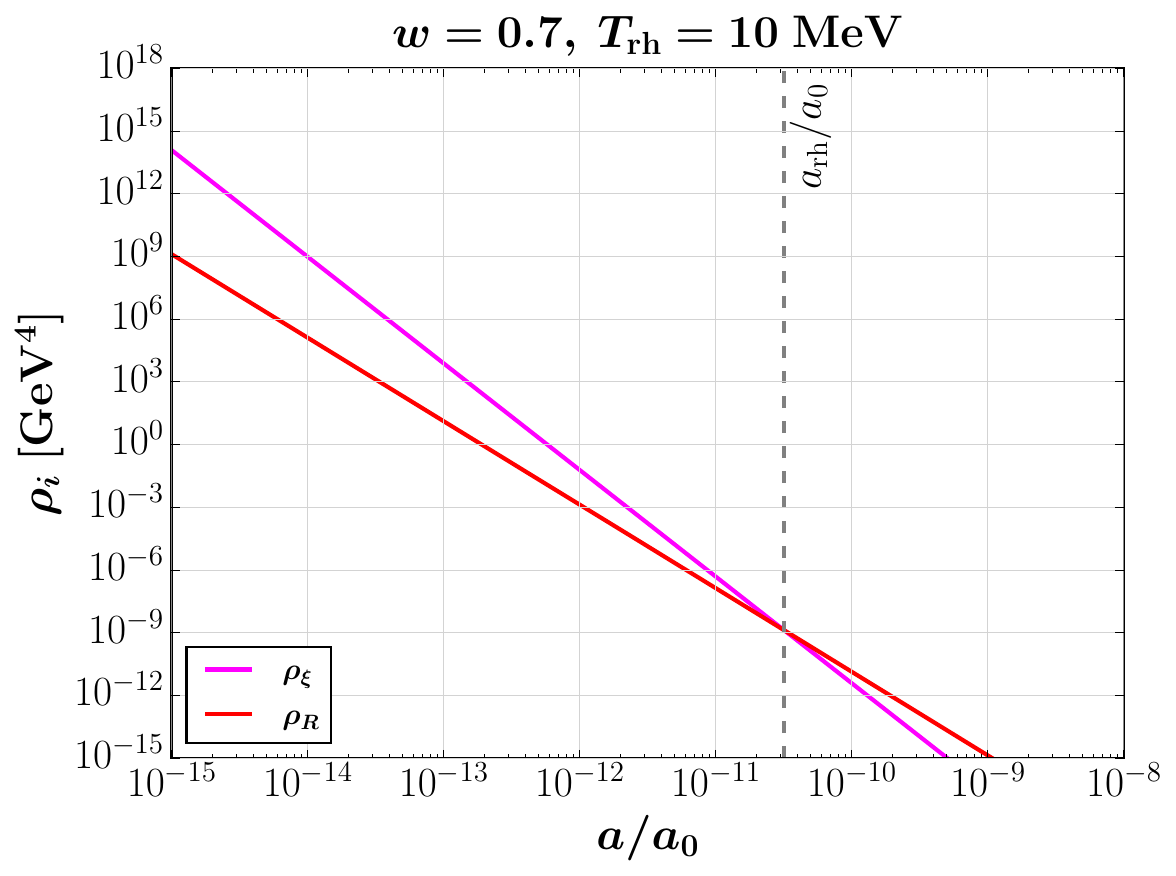}
\caption{\label{fig:EnergyDen} Evolution of the energy densities for radiation and fluid-$\xi$ as a function of scale factor for two different EoS $w$. The vertical dashed lines represent the corresponding scale factors at the reheating time.  Here, we take $m_{\psi}=1 \ \rm{GeV}$, $y_\psi = 4 \times 10^{-2}$, $ m_{h_2} = 10 \ \rm{MeV}$ and $T_{\rm rh}= \ 10 \ \rm{MeV}$.} 
\end{figure}

When $w > 1/3$, the energy density of the $\xi$-fluid, $\rho_\xi$, redshifts faster than radiation, where $\rho_R \propto a^{-4}$. The corresponding evolution of the energy densities for different values of the EoS $w$ is shown in Fig.~\ref{fig:EnergyDen}. Although the $\xi$-fluid may dominate the energy budget at early times, radiation inevitably overtakes it as the Universe expands. This triggers a natural transition back to the standard radiation-dominated era at a specific temperature, denoted as the reheating temperature $T_{\rm rh}$. To preserve the successful predictions of Big Bang Nucleosynthesis regarding the primordial abundances of light elements, this transition must complete before the BBN epoch, imposing a firm lower bound of $T_{\rm rh} \gtrsim 4$ MeV~\cite{Hannestad:2004px, Hasegawa:2019jsa}.

To computationally track the freeze-out dynamics, it is highly advantageous to transition to dimensionless variables. We define the comoving yield $Y_\psi = n_\psi / s$, where $s$ is the total entropy density of the Universe, and the dimensionless inverse temperature variable $x = m_\psi / T$. Applying these transformations to the dark matter number density equation allows it to be written as:\begin{equation}
	\frac{dY_\psi}{dx} = - \frac{s \langle \sigma v \rangle}{Hx} \left( Y_\psi^2 - Y_{\psi, \rm{eq}}^2 \right)
\end{equation}
Substituting the enhanced Hubble rate for the non-standard era into this framework, the pre-factor regulating the annihilation efficiency becomes significantly attenuated compared to the SM:
\begin{equation}\frac{dY_\psi}{dx} = - \sqrt{\frac{\pi g_*(T)}{45}} \frac{M_p m_\psi \sigma_1}{x^2} \left( \frac{T_{\rm rh}}{m_\psi/x} \right)^{\frac{3w-1}{2}} \left( Y_\psi^2 - Y_{\psi, \rm{eq}}^2 \right)
\end{equation}

Because the Hubble expansion $H$ is significantly larger at a given temperature $T$ than in the SM, the decoupling condition $\Gamma \simeq H$ is satisfied at a smaller value of $x$ (meaning freeze-out occurs at an earlier time and a higher temperature $T_f$). Since the equilibrium yield $Y_{\rm eq}$ drops exponentially as $e^{-x}$ for non-relativistic particles, freezing out at a smaller $x$ means the dark matter decouples while its abundance is much higher. Consequently, the final asymptotic yield $Y_\infty$ is vastly enhanced compared to the radiation-dominated case.

\begin{figure}[tbp]
\centering 
\includegraphics[width=.49\textwidth]{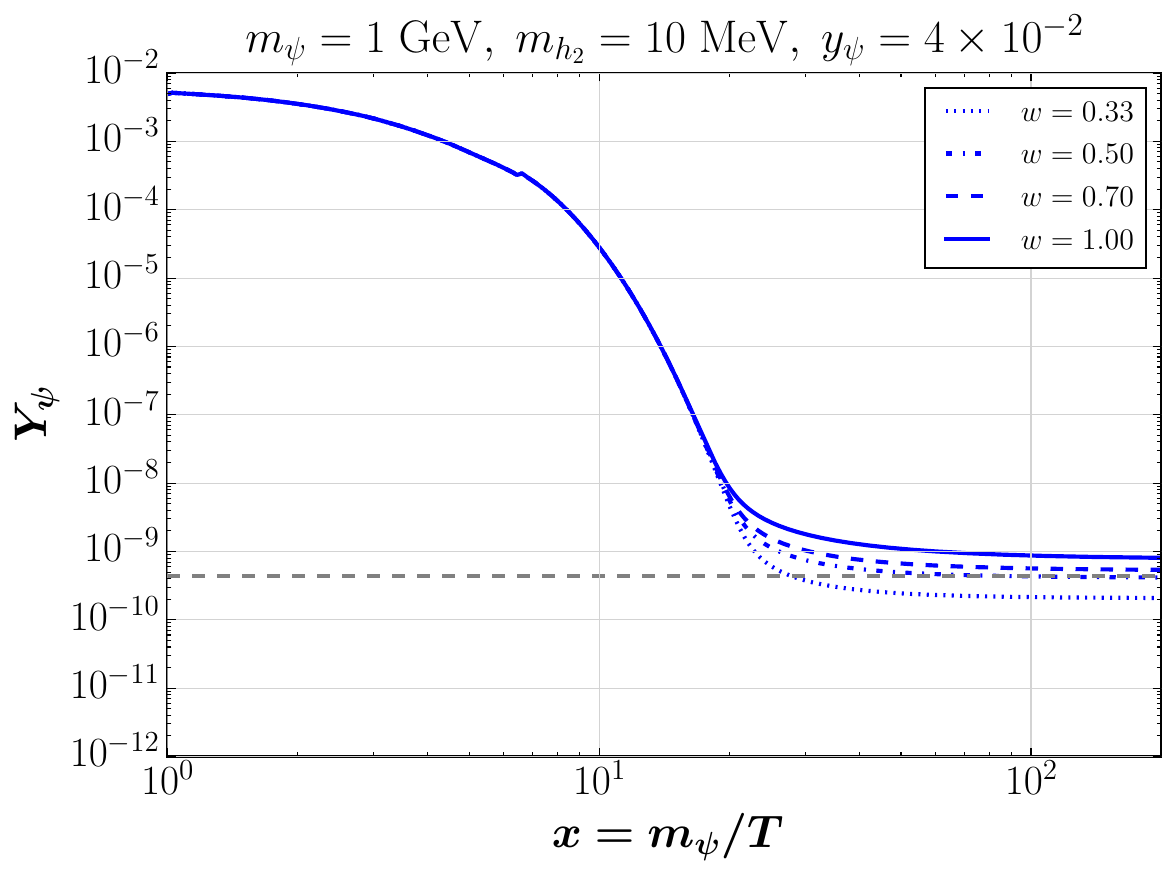}
\hfill
\includegraphics[width=.49\textwidth]{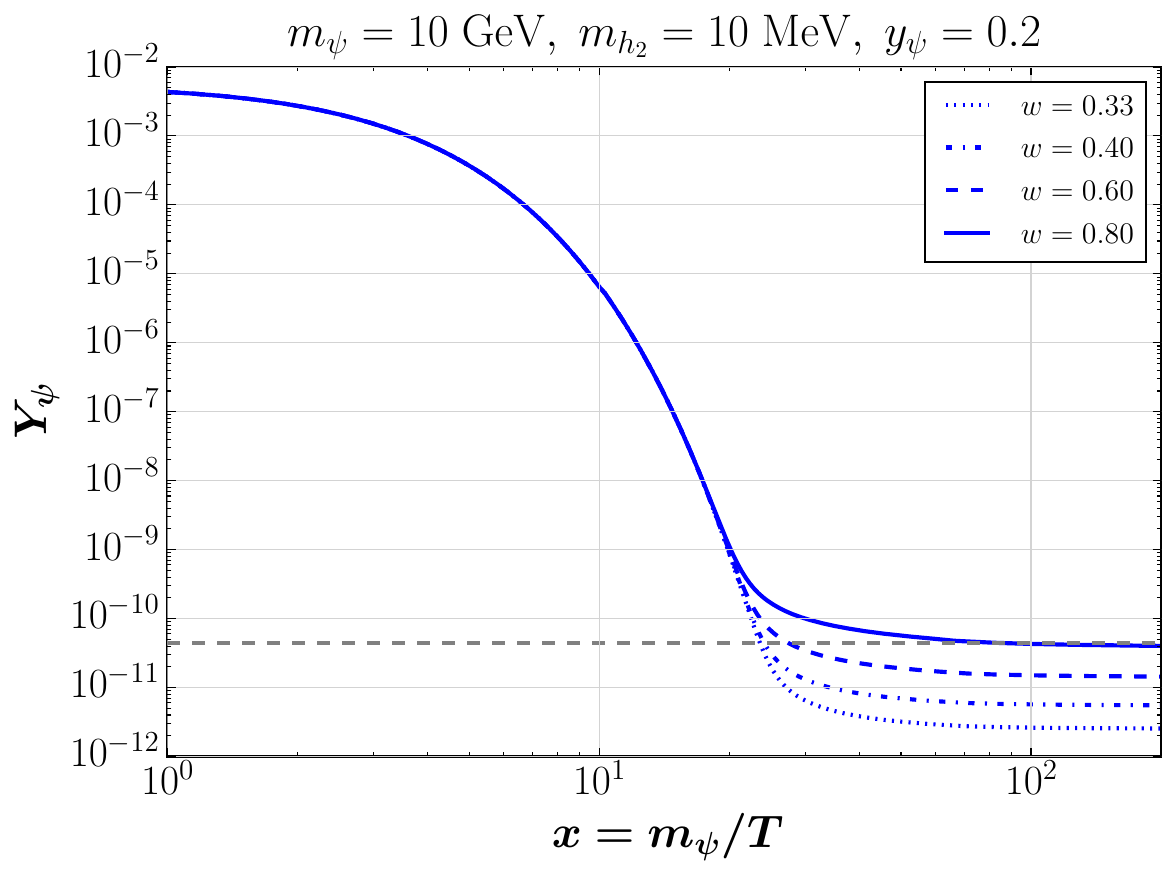}
\caption{DM yield evolution for different EoS $w$ with two different dark matter masses $m_{\psi}=1 \ \rm{GeV}$ and $10 \ \rm{GeV}$, {and $y_\psi=4\times 10^{-2}$ and $0.2$ respectively}. The dashed horizontal line represents the corresponding correct relic abundance. }
\label{fig:YDM}
\end{figure}

In Fig.~\ref{fig:YDM}, we show the numerical solution of the Boltzmann equation for dark matter yield with different values of the EoS $w$. When $w$ increases, DM freeze-out happens earlier and consequently the final yield is enhanced compared to the radiation-dominated case ($w=1/3$).

\begin{figure}[tbp]
\centering 
\includegraphics[width=.7\textwidth]{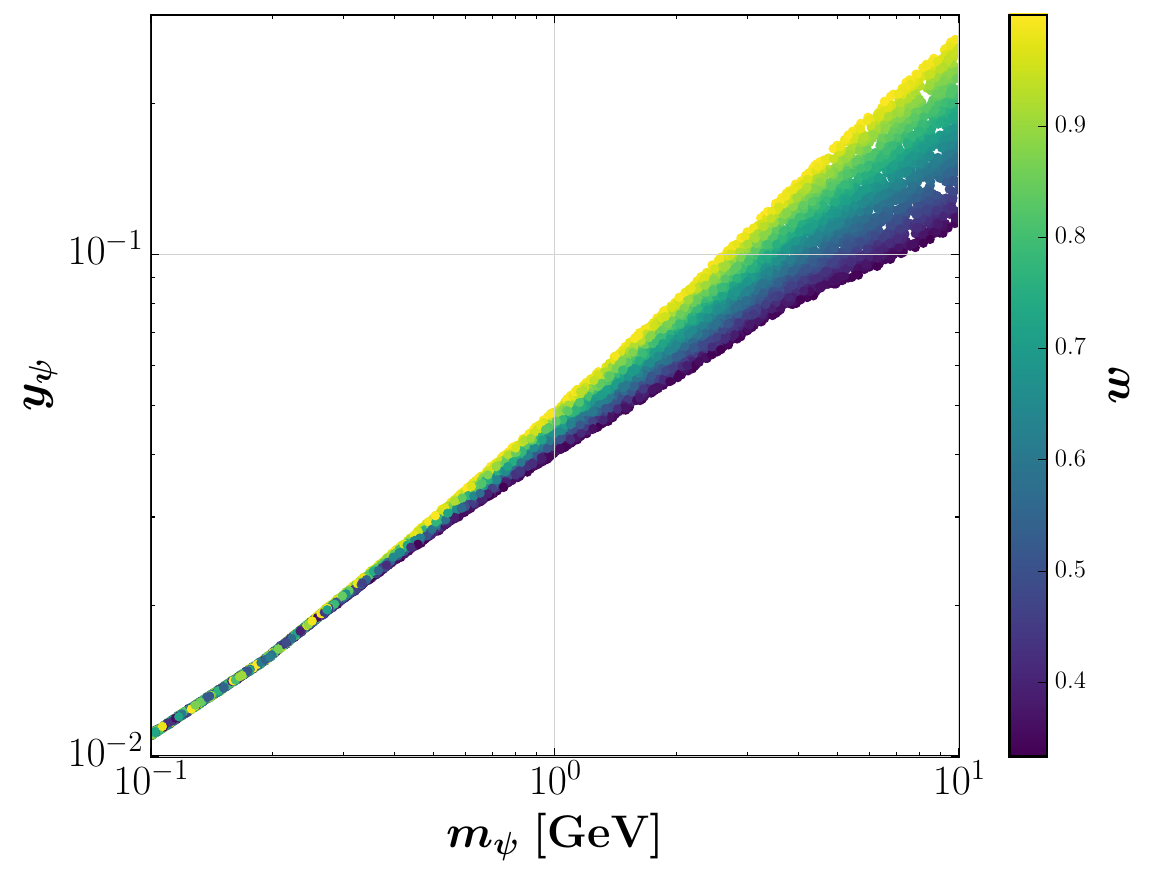}
\caption{\label{fig:contourYDM} Parameter space in the {$(m_\psi, y_\psi)$} plane consistent with the correct DM relic abundance, $\Omega_{\rm DM} h^2 = 0.1200 \pm 0.0012$ at 68\% C.L~\cite{Planck:2018vyg}. The color scale illustrates the EoS parameter.}
\end{figure}

In Fig.~\ref{fig:contourYDM}, we illustrate the parameter space in the {$(m_\psi, y_\psi)$} plane that yields the correct DM relic abundance, while varying the EoS $w$ in the range of $[1/3, 1]$. The mediator mass and reheating temperature are fixed at $m_{h_2}=10 \ \mathrm{MeV}$ and $T_{\rm rh}=10 \ \mathrm{MeV}$, respectively. We find that, for sub-GeV DM mass scale, achieving correct DM relic abundance requires a larger coupling $y_\psi$ during $\xi$-fluid dominated epoch, i.e, for  $w>1/3$ than in the standard RD case. Moreover, one can clearly realize that the dependence of the final DM relic abundance on the EoS parameter is negligible for  relatively light DM masses, namely {$m_\psi \lesssim 300 \ \rm{MeV}$.} This behaviour indicates that, in this mass range, DM freeze-out happens at essentially the same cosmological epoch, making the relic abundance largely insensitive to the value of $w$.   {This behavior is directly tied to the relationship between the dark matter freeze-out temperature $T_f$ and the chosen reheating temperature $T_{\rm rh}$. Because dark matter typically freezes out at $T_f \sim m_\psi / x_f$ (where $x_f \sim 15 - 20$), particles with $m_\psi \lesssim 300 \ \mathrm{MeV}$ decouple at temperatures near or below $T_{\rm rh} = 10 \ \mathrm{MeV}$. At these temperatures, the Universe has already transitioned (or is actively transitioning) out of the $\xi$-fluid epoch and back into the standard radiation-dominated era. Consequently, the decoupling dynamics are primarily governed by the standard radiation Hubble rate, making the relic abundance largely insensitive to the exact value of the non-standard EoS parameter $w$ that governed the earlier universe. Crucially, for the parameter space where freeze-out occurs prior to this transition, the requirement of an enhanced Yukawa coupling $y_\psi$ has profound phenomenological implications. Because the DM-nucleon scattering cross-section is driven by this same coupling, the non-standard cosmological history directly predicts an amplified scattering rate. This enhancement translates to a significantly higher number of expected events in direct detection experiments, elevating the overall probability of discovery, a synergistic link between the early-universe EoS and observational prospects that forms the central highlight of this work.}

\subsection{Evading the CMB and Indirect Search Constraints}
The $p$-wave velocity suppression of the annihilation cross-section is instrumental for the survival of the model. The Planck collaboration has mapped the CMB with unprecedented precision, placing stringent constraints on any exotic energy injection into the intergalactic medium during the recombination epoch ($z \sim 1100$). Annihilating dark matter can inject high-energy SM particles (into which the $h_2$ mediators ultimately decay) that ionize neutral hydrogen, increasing the residual free electron fraction and modifying the CMB spectra~\cite{Slatyer:2015jla, Baer:2014eja, Madhavacheril:2013cna}.

The rate of energy injection per unit volume is given by:
	\begin{equation}
		\left( \frac{dE}{dt} \right)_{\rm{inj}} = f(z) \rho_{\rm{DM}}^2 \frac{\langle \sigma v \rangle}{m_\psi},
	\end{equation}
    where $f(z)$ is an efficiency factor representing the fraction of annihilation energy deposited into the plasma.
For standard $s$-wave annihilating thermal relics, $\langle \sigma v \rangle \simeq 3 \times 10^{-26} \text{ cm}^3/\text{s}$, which is a constant. For light DMs, this constant cross-section injects far too much energy, strongly violating Planck limits. However, for the $p$-wave model, the cross-section depends on the velocity squared. Thus, by the time of recombination, the dark matter has cooled significantly due to the expansion of the Universe. 
Because the annihilation rate scales as $v^2$, the cross-section during recombination drops substantially relative to its value during freeze-out.

 {Furthermore, this same $p$-wave velocity suppression naturally evades present-day indirect search constraints. Indirect detection experiments (such as Fermi-LAT or H.E.S.S.) search for anomalous fluxes of gamma-rays or other cosmic rays originating from regions of high dark matter density, such as the Galactic Center or dwarf spheroidal galaxies~\cite{Elor:2015bho, Profumo:2017obk}. In these local astrophysical environments, the dark matter halo is cold and highly non-relativistic, characterized by typical velocities of $v \sim 10^{-3} c$ in the Milky Way and $v \sim 10^{-4} c$ in dwarf spheroidals. Consequently, the late-time annihilation cross-section is severely suppressed by factors of $10^{-6}$ to $10^{-8}$ compared to its value at freeze-out. This enormous suppression guarantees that the predicted fluxes of SM decay products fall orders of magnitude below the sensitivity thresholds of current and forthcoming indirect detection observatories.}

 {Therefore, the $p$-wave nature of the vector-like fermion dark matter cleanly and robustly escapes both the stringent early-Universe CMB bounds and late-Universe indirect detection limits, rendering the scalar portal a uniquely viable scenario.}

\subsection{Constraints from the Scalar Sector}
In addition to relic-density, direct-detection, and other requirements discussed so far, the scenario is subject to several cosmological and collider bounds that further shape the viable parameter space. In this section, we discuss the BBN constraint on the lifetime of $h_2$, Higgs invisible decay constraints~\cite{ATLAS:2022yvh}, the limits from supernova~\cite{Balaji:2022noj} and collider searches~\cite{Dev:2017dui, Egana-Ugrinovic:2019wzj, Dev:2019hho}. 
The light scalar $h_2$ can decay into the SM fermions after mixing with the SM Higgs boson. The partial decay width into a fermion pair is given as
\begin{eqnarray}
\Gamma_{h_2\rightarrow ff}=\sin^2\theta\left(\frac{m_f}{v_H}\right)^2\frac{m_{h_2}}{8\pi}\bigg(1-\frac{4m_f^2}{m^2_{h_2}}\bigg)^{\frac{3}{2}},
\end{eqnarray}
where $m_f$ is the mass of the fermion, $\sin\theta$ is the mixing angle between the SM Higgs and scalar $\phi$, $v_H$ is the VEV of SM Higgs. The singlet scalar must decay before the onset of BBN; otherwise, its late-time decay would disrupt the successful predictions of primordial nucleosynthesis.

\begin{figure}[tbp]
\centering 
\includegraphics[width=.7\textwidth]{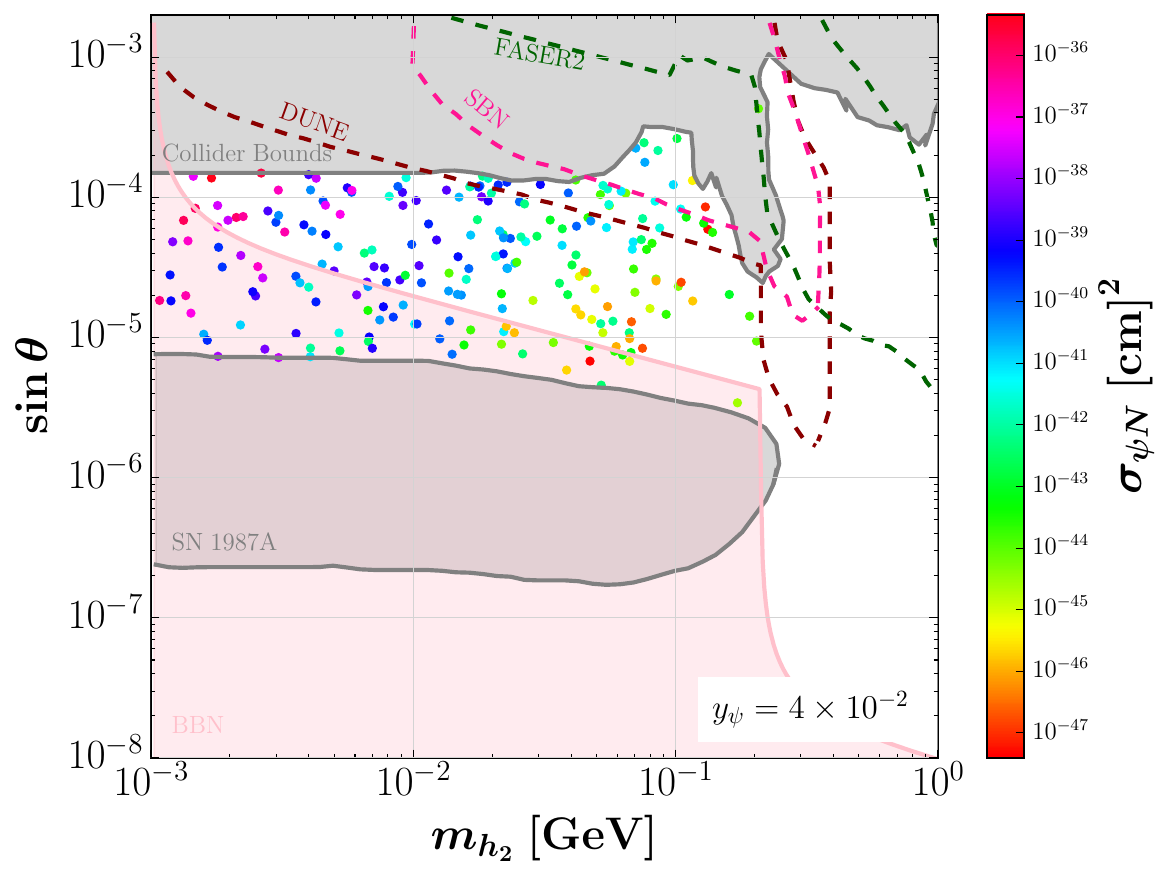}
\caption{\label{fig:sin_mh2} Parameter space in the $\sin \theta-m_{h_2}$ plane consistent with the current direct DM detection constraints from CRESST-III~\cite{CRESST:2019jnq}, DarkSide-50~\cite{DarkSide-50:2022qzh} and the neutrino floor~\cite{Hertel:2018aal}. The color scale illustrates the corresponding SI dark matter-nucleon elastic scattering cross section for the DM masses in the range $m_{\psi} \in [100 \ \rm{MeV}, \ 10 \ \rm{GeV} ]$ with the Yukawa coupling fixed at $y_\psi = 4\times 10^{-2}$. The grey shaded regions represent the existing collider bounds from Ref.~\cite{Batell:2022dpx} and the supernova constraint labeled by SN 1987A~\cite{Balaji:2022noj}, respectively. The light pink shaded region is excluded due to $\tau_{h_2} > \tau_{\rm BBN}$.  The dashed dark red, pink and dark green lines indicate the projected sensitivities of DUNE~\cite{Berryman:2019dme}, SBN~\cite{Batell:2019nwo} and FASER 2~\cite{Anchordoqui:2021ghd}.  }
\end{figure}

In Fig.~\ref{fig:sin_mh2}, we show these existing constraints together with the future sensitivities of DUNE \cite{Berryman:2019dme}, SBN \cite{Batell:2019nwo}, and FASER 2 \cite{Anchordoqui:2021ghd} {in the $m_{h_2}-\sin\theta$ plane.} The parameter space shown is consistent with the current direct DM detection constraints from CRESST-III~\cite{CRESST:2019jnq} and DarkSide-50~\cite{DarkSide-50:2022qzh} and lies above the neutrino floor~\cite{Hertel:2018aal}.
The color scale illustrates the corresponding  SI dark matter-nucleon elastic scattering cross section for the DM masses in the range $m_{\psi} \in [100 \ \rm{MeV}, \ 10 \ \rm{GeV} ]$ with the Yukawa coupling fixed at $y_\psi = 4\times 10^{-2}$. The benchmark value of $\sin\theta$ value is consistent with all current experimental and cosmological constraints while remaining within the projected reach of future experiments. 

\section{Stochastic Gravitational Wave Background in Non-standard Cosmology}\label{sec:gw}

Assuming a nearly scale-invariant primordial GW spectrum, the late-time GW spectrum acquires a characteristic shape, with any deviations serving as potential signatures of a non-standard thermal history. Since direct probes of the dynamics of Universe prior to BBN are limited, analyzing such deviations offers a valuable indirect window into early universe physics. In particular, the shape of the late-time stochastic gravitational wave background can reflect modifications to the expansion history, where the spectral density typically exhibiting a high-frequency plateau can either be enhanced or suppressed depending on the equation of state during the pre-BBN era.

Therefore, our analysis establishes a correlation between the non-standard expansion history of the early Universe, characterized by the fluid-$\xi$, its imprint on the stochastic GW background, and the DM phenomenology. In particular, the same parameter space that reproduces the correct DM relic abundance also gives rise to distinctive GW signatures and potentially observable direct detection signals.

The gravitational waves are defined by the transverse-traceless (TT) part of the metric perturbation in a spatially flat background
\dis{
ds^2 = a^2(\tau) [-d\tau^2 + (\delta_{ij} + h_{ij}) dx^i dx^j],
}
where $\tau$ is conformal time and the tensor perturbation $h_{ij}$ satisfy TT conditions, $\partial^i h_{ij} = h_i^i=0$. From the linear Einstein field equation $\delta G_{ij} = 8\pi G \delta T_{ij}$, one can find the equation of motion for GW
\dis{
\ddot{h}_{ij}(t, \boldsymbol{x}) + 3 H \dot{h}_{ij}(t,\boldsymbol{x}) - \frac{\bigtriangledown^2}{a^2}h_{ij}(t,\boldsymbol{x})= 16 \pi G \Pi^{\rm{TT}}_{ij}(t,\boldsymbol{x}),
}
where $\Pi^{\rm{TT}}_{ij}(t, \boldsymbol{x})$ is the transverse and traceless part of the anisotropic part of the stress tensor, defined by $T_{ij} = p \ g_{ij} + a^2 \Pi_{ij}$, where $\Pi_{ij}=0$ for a perfect fluid and the Laplacian $\bigtriangledown^2$ has been replaced by the comoving wavenumber $-k^2$ in the third term. 
 The Fourier modes of $h_{ij}$ can be decomposed into the two polarization states $\lambda=+,\times$, as
\dis{
h_{ij}(t, \boldsymbol{x}) = \sum_{\lambda=+,\times} \int \frac{d^3 \boldsymbol{k}}{ (2\pi)^{3/2}} h_\lambda(t, \boldsymbol{k}) e^{i \boldsymbol{k} \cdot \boldsymbol{x}} \epsilon_{ij}^\lambda(\hat{\boldsymbol{k}}),
}
where $\epsilon_{ij}^\lambda(\hat{\boldsymbol{k}})$ are the two polarization tensors, normalized as $\sum_{i,j} \epsilon_{ij}^\lambda(\hat{\boldsymbol{k}}) \epsilon_{ij}^{\lambda'}(\hat{\boldsymbol{k}}) = 2 \delta_{\lambda \lambda'}$.
Therefore, using conformal time derivative $('\equiv \frac{\partial}{\partial \tau})$ with {$d\tau = a^{-1} dt$,} we obtain the equation for gravitational wave in a perfect fluid as
\dis{\label{eq:GWeom_conformal}
h_{ij}''(\tau, \boldsymbol{k}) + \frac{2 a'}{a} h_{ij}'(\tau, \boldsymbol{k}) + k^2 h_{ij}(\tau, \boldsymbol{k})=0.
}
General solution at any time can be represented by 
\dis{\label{eq:hijSol}
h_\lambda(\tau, \boldsymbol{k}) = h_{\lambda, \rm{prim}} ({k}) \mathcal{T}(\tau, {k}),
}
where $h_{\lambda, \rm{prim}} ({k})$ is the amplitude of the primordial tensor perturbations and $\mathcal{T}(\tau, {k})$ is the transfer function, reflecting the evolution of GWs after the relevant GW modes cross the horizon. As long as a mode remains outside the horizon, the corresponding perturbation does not vary with time, so that the transfer function is very well approximated by $\mathcal{T}(\tau, {k})=1$. 

The energy density of gravitational waves is defined by~\cite{Turner:1993vb, Chongchitnan:2006pe, Nakayama:2008wy, Kuroyanagi:2011fy}
\begin{eqnarray}\label{eq:EnerDenGW}
    \rho_{\rm GW}(\tau) &\equiv& \int_0^\infty \frac{dk}{k} \frac{d \rho_{\rm GW}}{d \ln k} =  \frac{\langle h'_\lambda(\tau, \boldsymbol{k}) h'_\lambda(\tau, \boldsymbol{k})\rangle}{32 \pi G a^2(\tau)} \nonumber \\
    &=& \frac{2}{32 \pi G a^2(\tau)}  \int \frac{d^3 k d^3 k'}{(2 \pi)^3} \langle h'_\lambda(\tau, \boldsymbol{k}) h'_\lambda(\tau, \boldsymbol{k}')\rangle e^{i (\boldsymbol{k} + \boldsymbol{k}') \cdot \boldsymbol{x}} \nonumber \\
    &=& \frac{1}{16 \pi G a^2(\tau)}  \int \frac{d^3 k}{(2 \pi)^3} |h'_\lambda(\tau, \boldsymbol{k})|^2 = \frac{1}{32 \pi^3 G a^2(\tau)} k^2 dk |h'_\lambda(\tau, \boldsymbol{k})|^2,
\end{eqnarray}
where we have used $\langle h'_\lambda(\tau, \boldsymbol{k}) h'_\lambda(\tau, \boldsymbol{k'})\rangle = \delta^{(3)}(\boldsymbol{k} + \boldsymbol{k}') |h'_\lambda(\tau, \boldsymbol{k})|^2$ in the third line. 
Consequently, using the Eq.~(\ref{eq:hijSol}), the differential energy density of GW finds
\dis{
\frac{d \rho_{\rm GW}}{d \ln k} = \frac{k^3}{32 \pi^3 G a^2(\tau)} |h_{\lambda, \rm{prim}}(k)|^2 [\mathcal{T}'(\tau, k)]^2.
}
The spectrum of GWs is described in terms of the fraction of their energy density per logarithmic frequency interval
\dis{\label{eq:OmGW}
\Omega_{\rm GW} (\tau, k) = \frac{1}{\rho_{\rm cr}} \frac{d\rho_{\rm GW}(\tau, k)}{d \ln k} = \frac{1}{12 a^2(\tau) H^2(\tau)} \mathcal{P}^{\rm prim}_T(k) [\mathcal{T}'(\tau, k)]^2,
}
where $\rho_{\rm cr} = 3H^2/8\pi G$ is the critical energy density of the universe. 
When the modes are deep inside the horizon ($k\gg \mathcal{H}$), the transfer function is usually given by Bessel functions as $\mathcal{T}(x) = \frac{1}{x^n} [A j_n(x) + B y_n(x)]$ and $\frac{d}{d \tau} \mathcal{T}(x) = -\frac{k}{x^n} [A j_{n+1}(x) + B y_{n+1}(x)]$ with $x=k \tau$. Hence, it is customary to approximate $[\mathcal{T}'(\tau, k)]^2 \simeq k^2 \mathcal{T}^2(\tau, k)$~\cite{Watanabe:2006qe, Caprini:2018mtu}.

 In order to facilitate the comparison between the inflationary predictions and the observational data, it is instructive to express the inflationary tensor power spectra on super-Hubble scales as power laws around a pivot scale $k_*$, such as 
\dis{\label{eq:Pt_prim_obs}
\mathcal{P}^{\rm prim}_T(k) = r \mathcal{P}_\xi(k_{*}) \left(\frac{k}{k_{*}}\right)^{n_T},
}
where $r$ is the tensor-to-scalar ratio, % {$n_T\simeq - r/8$ -> $n_T$} 
$n_T$ is the spectral index and $\mathcal{P}_\xi(k_{*})$ is the power spectrum for the scalar perturbation, measured as $r<0.036$ and $\mathcal{P}_\xi=2.1\times 10^{-9}$, respectively, at the CMB pivot scale $k_{*} = 0.05 \rm{Mpc}^{-1}$~\cite{Planck:2018jri}. Since $n_T<0$, the primordial tensor spectrum is nearly scale invariant with a slightly red tilt.
\subsection{Enhanced Gravitational Wave Signature}
When $k\gg \mathcal{H}$, the solution to the GW equation of motion, $h_\lambda(\tau, \boldsymbol{k}) $, oscillates with a decaying envelope and can be approximated by~\cite{Boyle:2005se, Caprini:2018mtu} 
\dis{
h_\lambda(\tau, \boldsymbol{k}) \simeq h_{\lambda, \rm{prim}} ({k}) \frac{a_k}{a(\tau)} \cos{[k (\tau-\tau_k) + \phi_k]}\,,
}
where $\phi_k$ is phase shift of the oscillation, and $\tau_k$ is the conformal time at the horizon crossing. Consequently, from the Eq.~(\ref{eq:hijSol}), the transfer function at horizon entry during an earlier epoch is related to its present value by
\dis{
\mathcal{T}(\tau_0, k) \simeq \cos{[k (\tau_0-\tau_k) + \phi_k]} \frac{a_k}{a_0} \simeq \frac{1}{\sqrt{2}} \frac{a_k}{a_0}, 
}
where the factor $1/\sqrt{2}$ comes from averaging over the oscillatory transfer function.
Using Eq.~(\ref{eq:OmGW}), the spectral GW energy density at the present is written as 
\dis{\label{eq:OmGW_scalFac}
\Omega_{\rm GW} (k) \equiv \Omega_{\rm GW} (\tau_0, k) \simeq \frac{1}{24} \left(\frac{k}{a_0 H_0}\right)^2 \mathcal{P}_T^{\rm prim}(k) \left(\frac{a_k}{a_0}\right)^2\,,
}
where $a_k$ is the scale factor at the horizon crossing, $k=a_k H_k$ is related to the observed GW frequency $f=k/(2 \pi a_0)$ and the Hubble parameter today $H_0 = 100 \ h \ \rm{km \ s^{-1} \ Mpc^{-1}} \simeq 2.18 \times 10^{-18} \ \rm{s^{-1}}$ with $h=0.674$. 

For scales which enter the horizon during radiation dominated era after reheating, i.e. $ a_{\rm rh} < a_{k} < a_{\rm eq}$, where $a_{\rm rh}$ and $a_{\rm eq}$ denote the scale factors at the end of reheating and at the matter-radiation equality, respectively, Eq.~(\ref{eq:OmGW_scalFac}) becomes 
\dis{\label{eq:OmGWRD}
\Omega_{\rm GW}^{R}(k)
\simeq 
\frac{\Omega_{R, 0}}{24} \mathcal{P}_T^{\rm prim}(k) \left(\frac{g_*(T)}{g_{*}(T_0)}\right)\left(\frac{g_{*,s}(T_0)}{g_{*,s}(T)}\right)^{4/3}
\,,
}
where $T$ is the temperature of radiation bath at horizon entry. To obtain this expression, we employed the following relation, which is derived from entropy conservation combined with the standard temperature dependence of the radiation energy density:

\dis{
\left( \frac{H_k}{H_0}\right)^2 \simeq  \Omega_{R, 0} \left(\frac{g_*(T)}{g_{*}(T_0)}\right)\left(\frac{g_{*,s}(T_0)}{g_{*,s}(T)}\right)^{4/3} \left(\frac{a_0}{a_k}\right)^4\,. 
}
The gravitational wave energy density corresponding to the scale $k_{\rm rh}$, which enters the horizon at the end of reheating, can be estimated as:
\dis{\label{eq:OmGWRD_RH}
\Omega_{\rm GW}^{R}(k_{\rm rh}) \simeq 2.4 \times 10^{-16} \left(\frac{k}{k_{*}}\right)^{n_T},
}
where we have used the present radiation energy density fraction $\Omega_{R, 0} \simeq 9.2 \times 10^{-5}$.
For the numerical estimation, the effective relativistic degrees of freedom at reheating and at present were taken to be $g_{*,s}(T_{\rm rh}) \simeq g_*(T_{\rm rh}) \simeq 10.75$, $g_{*}(T_0) \simeq 3.36$, and $g_{*,s}(T_0)\simeq 3.91$.

For scales that enter the horizon during the fluid-$\xi$ dominated epoch (i.e., $a_{\rm end} < a_k < a_{\rm rh}$, where $a_{\rm end}$ represents the scale factor at the end of inflation), Eq.~(\ref{eq:OmGW_scalFac}) can be expressed as:

\dis{\label{eq:OmGWStiff}
\Omega_{\rm GW}^{\xi} (k) &\simeq \frac{\mathcal{P}_T^{\rm prim}(k)}{24} \left(\frac{H_k}{H_0}\right)^2  \left(\frac{a_k}{a_0}\right)^4 \simeq 
\frac{\mathcal{P}_T^{\rm prim}(k)}{24} \left(\frac{H_k}{H_{\rm rh}}\right)^2 \left(\frac{H_{\rm rh}}{H_0}\right)^2   \left(\frac{a_k}{a_{\rm rh}}\right)^4 \left(\frac{a_{\rm rh}}{a_0}\right)^4 \\
&\simeq  \Omega_{\rm GW}^{R}(k_{\rm rh}) \left(\frac{k}{k_{\rm rh}}\right)^{n_T} \left(\frac{H_k}{H_{\rm rh}}\right)^2  \left(\frac{a_k}{a_{\rm rh}}\right)^4 
\simeq \Omega_{\rm GW}^{R}(k_{\rm rh}) \left(\frac{k}{k_{\rm rh}}\right)^{n_T} \left(\frac{a_k}{a_{\rm rh}}\right)^{1-3w} \,,
}

Here, we have applied the sudden transition approximation between the $\xi$-dominated and radiation-dominated eras. To arrive at the final expression, we utilized Eq.~(\ref{eq:OmGWRD}) for the reheating wavenumber $k_{\rm rh}$, along with the Hubble expansion scaling relation $(H_k/H_{\rm rh})^2 \propto \rho_k/\rho_{\rm rh} \propto (a_k/a_{\rm rh})^{-3(1+w)}$.

 During this fluid-$\xi$ dominated epoch, the GW frequency corresponding to the horizon entry redshifts into the present as
\dis{\label{eq:f_stiff}
f &= \frac{H_k}{2 \pi} \frac{a_k}{a_0} \simeq \frac{\rho^{1/2}_{\rm rh}}{2 \sqrt{3} \pi M_p} \frac{a_k}{a_0} \left(\frac{a_k}{a_{\rm rh}}\right)^{\frac{-3(1+w)}{2}} \simeq \frac{\rho^{1/2}_{\rm rh}}{2 \sqrt{3} \pi M_p} \frac{a_k}{a_{\rm rh}} \frac{a_{\rm rh}}{a_0}  \left(\frac{a_k}{a_{\rm rh}}\right)^{\frac{-3(1+w)}{2}} \\
&\simeq \frac{1}{2} \sqrt{\frac{g_{*}(T_{\rm rh})}{90}} \frac{T_{\rm rh} T_0}{M_p} \left(\frac{g_{*,s}(T_0)}{g_{*,s}(T_{\rm rh})}\right)^{1/3} \left(\frac{a_k}{a_{\rm rh}}\right)^{\frac{-(1+3w)}{2}} \simeq f_{\rm rh} \left(\frac{a_k}{a_{\rm rh}}\right)^{\frac{-(1+3w)}{2}}\,,
}
where $f_{\rm rh}$ is the frequency corresponding to reheating, which can be numerically estimated as $f_{\rm rh} \simeq 1.8 \times 10^{-10}$ for $T_{\rm rh} = 10 \ \rm{MeV}$. Using Eq.~(\ref{eq:f_stiff}), the GW spectrum during the fluid-$\xi$ dominated epoch can be written in terms of frequency $f$ as
\begin{align}
\label{eq:OmGW0_stiff}
    \Omega_{\rm GW}^{\xi} (f) \simeq \Omega_{\rm GW}^{R}(f_{\rm rh}) \left(\frac{f}{f_{\rm rh}}\right)^{n_T} \left(\frac{f}{f_{\rm rh}}\right)^{\frac{-2(1-3w)}{1+3w}}\,.
\end{align}

\begin{figure}[tbp]
\centering 
\includegraphics[width=.49\textwidth]{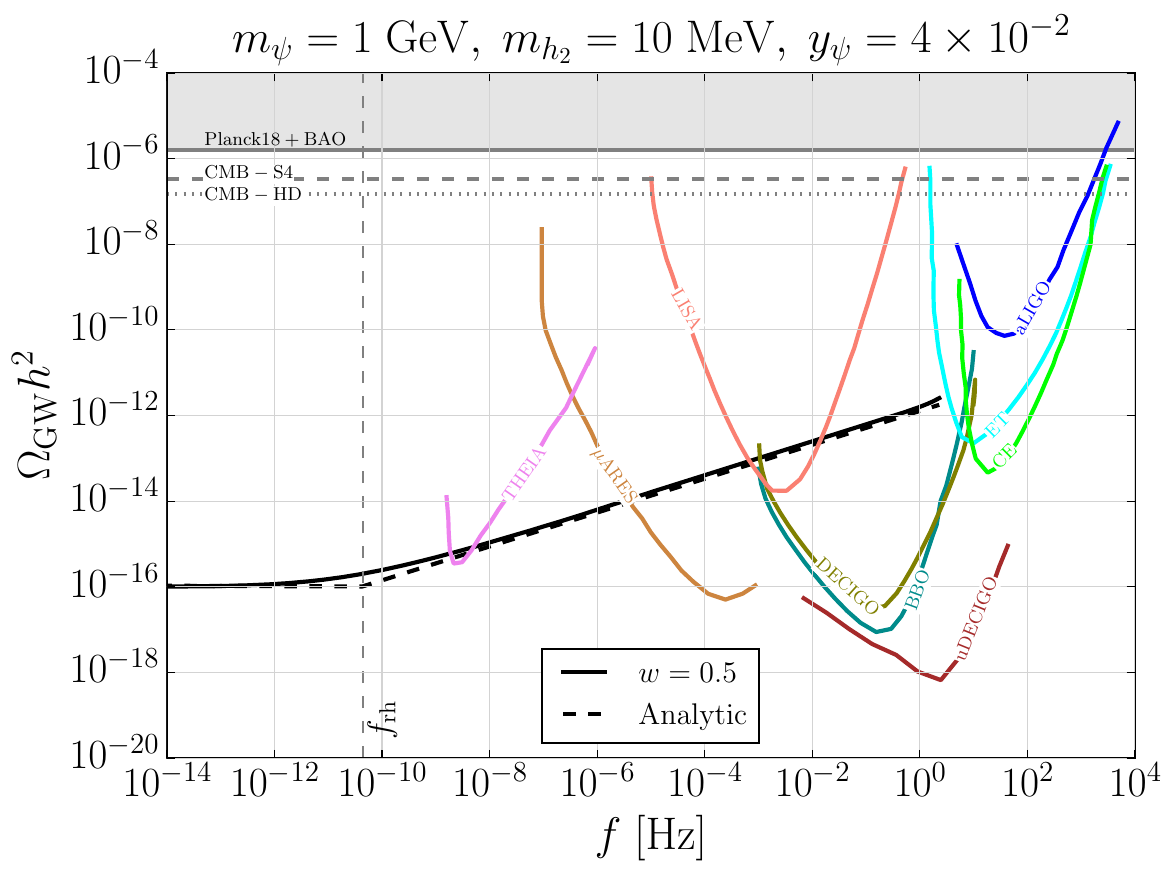}
\hfill
\includegraphics[width=.49\textwidth]{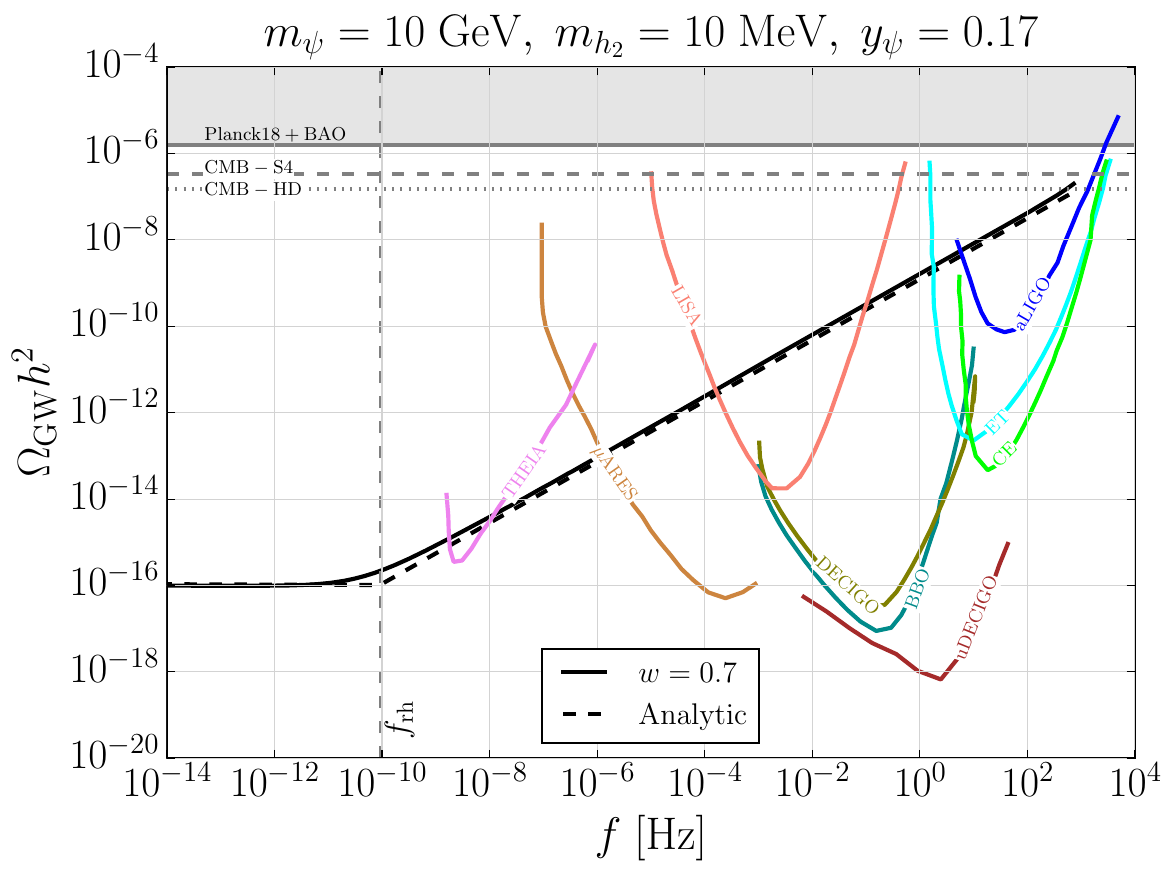}
\caption{
The present-day gravitational wave spectra as a function of observed frequency for two benchmark points: $m_\psi=1 \text{ GeV}$, $m_{h_2} =  10 \text{ MeV}$, and $y_\psi = 4 \times 10^{-2}$ with an equation of state $w=0.5$ (left panel); and $m_\psi=10 \text{ GeV}$, $m_{h_2} =  10 \text{ MeV}$, and $y_\psi = 0.17$ with a stiffer equation of state $w=0.7$ (right panel). The different values of the EoS parameter $w$ naturally result in distinct spectral tilts and different cutoff frequencies ($f_{\text{end}}$) for each scenario. In both panels, the dashed curves represent the analytical approximations derived in Eq.~(\ref{eq:OmGW0_stiff}), while the vertical gray dashed lines indicate the frequency corresponding to the end of reheating. The expected sensitivity curves of various operating and forthcoming GW observatories are superimposed, including ground-based interferometers (aLIGO~\cite{LIGOScientific:2014pky}, ET~\cite{Sathyaprakash:2012jk, ET:2019dnz}, CE~\cite{Reitze:2019iox}), space-based interferometers (LISA~\cite{LISA:2017pwj}, BBO~\cite{Crowder:2005nr, Corbin:2005ny}, DECIGO/uDECIGO~\cite{Seto:2001qf, Kudoh:2005as}, $\mu$-ARES~\cite{Sesana:2019vho}), the THEIA star survey recast~\cite{Garcia-Bellido:2021zgu}, and CMB polarization measurements (Planck 2018~\cite{Planck:2018vyg}, CMB-S4~\cite{CMB-S4:2016ple}, and CMB-HD~\cite{CMB-HD:2022bsz}).}
\label{fig:GW_spectrum}
\end{figure}

In order to calculate the maximum frequency, we need to specify the evolution of the energy density between the end of inflation and the onset of RD epoch, namely during the fluid-$\xi$ dominated epoch. The maximum frequency, $f_{\rm end}$, corresponding to the beginning of the fluid-$\xi$ dominated epoch, is given by
\begin{equation}
    f_{\rm end } = \frac{H(T_{i})}{2 \pi} \frac{a_{i}}{a_0}\,,
\end{equation}
where $H(T_i) \simeq  \rho_\xi(T_{i})^{1/2}/(\sqrt{3} M_p)$ is the Hubble rate at the beginning of the fluid-$\xi$ dominated epoch.
Using Eq.~(\ref{eq:rho_xi_Tin}) and the entropy conservation, the relation between the scale factors at the beginning of the fluid-$\xi$ dominated epoch and the present is found as
\dis{\label{eq:aend_a0}
\frac{a_i}{a_0} &\simeq \left(\frac{g_{*,s}(T_0)}{g_{*,s}(T_{i})}\right)^{1/3} \frac{T_0}{T_{i}} \simeq 
\left(\frac{g_{*,s}(T_0)}{g_{*,s}(T_{\rm rh})}\right)^{1/3} \frac{T_0}{T_{\rm rh}} \left(\frac{\pi^2/30 \ g_*(T_{\rm rh}) T_{\rm rh}^4 } {\rho_\xi(T_{i})}\right)^{\frac{1}{3(1+w)}}\,.
}
Consequently, $f_{\rm end}$ becomes 
\dis{
f_{\rm end} \simeq \frac{T_0 ~\rho_\xi^{\frac{1+3w}{6(1+w)}}(T_{i})}{2\sqrt{3} \pi M_p} \left(\frac{g_{*,s}(T_0)}{g_{*,s}(T_{\rm rh})}\right)^{1/3} \frac{[\pi^2/30 \ g_*(T_{\rm rh})]^{\frac{1}{3(1+w)}}}{T_{\rm rh}^{\frac{3w-1}{3(1+w)}}}\,,
}
where the present temperature is $T_0 \simeq 2.35 \times 10^{-13} \ \rm{GeV}$.
Throughout this work, we assume that $T_i$ is much larger than the DM mass and fix it to be $T_{i} = 10^6 \ \rm{GeV}$. In this case, for the reheating temperature of $T_{\rm rh} = 10 \ {\rm MeV}$ and $w=0.5$, we obtain $\rho_\xi(T_{i}) \simeq 1.1 \times 10^{30} \ \rm{GeV}$. As a result, the maximum frequency is found as
\dis{
f_{\rm end} \simeq 4.7 \ \rm{Hz}\,.
}
% {Why is this one in new line? }
Moreover, for a stiffer EoS with $w=0.7$, the corresponding maximum frequency increases to $f_{\rm end}\simeq 1.5 \times 10^3 \ \rm{Hz}$ with $\rho_\xi(T_{i}) \simeq 1.1 \times 10^{35} \ \rm{GeV}$.  
These analytic estimations are in good agreement with our numerical results, where the GW spectrum terminates at the corresponding cutoff frequencies, as shown in Fig.~\ref{fig:GW_spectrum}.

The enhanced stochastic GW background contributes to the total radiation energy budget of the Universe. Consequently, it is constrained by measurements of the effective number of relativistic degrees of freedom ($N_{\text{eff}}$), which is defined via the total radiation energy density:
\dis{\rho_{\text{rad}} = \rho_\gamma \left[1+\frac{7}{8}\left(\frac{4}{11}\right)^{4/3} N_{\text{eff}}\right],}
where $N_{\text{eff}} = N_{\text{eff}}^{\text{SM}} + \Delta N_{\text{eff}}$. Here, $N_{\text{eff}}^{\text{SM}}=3.046$~\cite{Mangano:2005cc} is the Standard Model prediction, and $\Delta N_{\text{eff}}$ encompasses any additional relativistic contributions, including that of the stochastic GW background.  

Using the present-day photon energy density, $\rho_\gamma = \frac{\pi^2}{15} T_0^4$, and approximating the present GW energy density as $\rho_{\text{GW}} = \rho_{\text{cr,0}} \int d \ln f \ \Omega_{\text{GW}}(f)$ (where the present critical energy density is $\rho_{\text{cr,0}} \simeq 8.15 \times 10^{-47} h^2 \ \text{GeV}^4$), we find:

\dis{\Delta N_{\text{eff}} = \frac{8}{7}\left(\frac{11}{4}\right)^{4/3} \frac{\rho_{\text{GW}}}{\rho_\gamma} \simeq \frac{120}{7 \pi^2} \left(\frac{11}{4}\right)^{4/3} \frac{\rho_{\text{cr, 0}}}{T_0^4}   \int d \ln f \ \Omega_{\text{GW}}(f)\,.}

Current cosmological observations place strict upper bounds on $\Delta N_{\text{eff}}$, which subsequently constrain the integrated GW spectrum. For instance, a joint analysis of Planck and BAO data requires $\Delta N_{\text{eff}} < 0.28$ at the 95$\%$ confidence level~\cite{Planck:2018vyg}. Translating this limit to the GW spectrum yields an upper bound on its integrated energy density:
\dis{\label{eq:OmGW_BBN}
{\Omega_{\rm GW}h^2=}\int d \ln f \ \Omega_{\text{GW}}(f) h^2 \simeq 5.6 \times 10^{-6} \Delta N_{\text{eff}} < 1.6\times 10^{-6}.}

Finally, using Eq.~(\ref{eq:OmGWRD}) and Eq.~(\ref{eq:OmGW0_stiff}), we can evaluate the maximum integrated GW relic abundance today, incorporating the enhanced contribution from the fluid-$\xi$ dominated epoch. This evaluates to:

\dis{
    \Omega_{\text{GW}} &= \int_{f_{\text{eq}}}^{f_{\text{end}}} d \ln f \left[ \Omega_{\text{GW}}^R(f)+\Omega_{\text{GW}}^\xi(f) \right] \\\nonumber     &\simeq \Omega_{\text{GW}}^R(f_{\text{rh}}) \frac{1+ 3 w}{6w-2} \left(\frac{f}{f_{\text{rh}}}\right)^{\frac{6 w-2}{3w+1}}\Bigg|_{f_{\text{rh}}}^{f_{\text{end}}}\simeq \left[ n_T - \frac{2(1-3w)}{1+3w}\right]^{-1}\Omega_{\text{GW}}^\xi(f_{\text{end}}) .   
}

For a benchmark scenario with $w=0.6$ and a cutoff frequency of $f_{\text{end}} \simeq 83.7 \text{ Hz}$, the corresponding maximum integrated GW energy density is $\Omega_{\text{GW}}(f_{\text{end}}) \simeq 2 \times 10^{-9}$, which comfortably satisfies the BBN constraint presented in Eq.~(\ref{eq:OmGW_BBN}). In contrast, for a stiffer equation of state with $w=0.8$ (where $f_{\text{end}} \simeq 2.4 \times 10^4 \text{ Hz}$), the integrated GW energy density surges to $\Omega_{\text{GW}}(f_{\text{end}}) \simeq 1.3 \times 10^{-4}$, thereby violating the BBN bound. Imposing the $\Delta N_{\text{eff}}$ constraint requires that the maximum frequency for this specific $w=0.8$ scenario be restricted to $f_{\text{end}} \lesssim 76 \text{ Hz}$. It should be emphasized that this upper bound on the cutoff frequency is highly sensitive to the underlying cosmological parameters, most notably the equation of state parameter $w$, as well as the initial and reheating temperatures ($T_i$ and $T_{\text{rh}}$), because these variables collectively govern the spectral tilt and the overall amplitude of the blue-shifted GW spectrum.

Future observations from CMB-S4~\cite{CMB-S4:2016ple} and CMB-HD~\cite{CMB-HD:2022bsz} are expected to tighten the constraint on $\Delta N_{\text{eff}}$ to $\lesssim 0.06$ and $\lesssim 0.027$, respectively. In this work, we evaluate the detectability of our predicted GW spectra by comparing them against these current and projected cosmological bounds, as well as the expected sensitivity curves of several forthcoming GW observatories. These include the Big Bang Observer (BBO)~\cite{Crowder:2005nr, Corbin:2005ny}, ultimate DECIGO (uDECIGO)~\cite{Seto:2001qf, Kudoh:2005as}, LISA~\cite{LISA:2017pwj}, $\mu$ARES~\cite{Sesana:2019vho}, THEIA~\cite{Garcia-Bellido:2021zgu}, Cosmic Explorer (CE)~\cite{Reitze:2019iox}, the Einstein Telescope (ET)~\cite{Sathyaprakash:2012jk, ET:2019dnz}, and Advanced LIGO (aLIGO)~\cite{LIGOScientific:2014pky}. The results of this comprehensive analysis are depicted in Fig.~\ref{fig:GW_spectrum}.

In Fig.~\ref{fig:GW_spectrum}, we present the stochastic gravitational wave spectrum as a function of frequency for  two benchmark scenarios that successfully reproduce the correct DM relic abundance (as previously identified in Fig.~\ref{fig:YDM}). The analytical estimations from Eq.~(\ref{eq:OmGW0_stiff}) are plotted as dashed curves and show excellent agreement with the numerical results. For modes entering the horizon prior to the end of reheating ($a \leq a_{\text{rh}}$, or $f \geq f_{\text{rh}}$), the enhanced expansion rate during the $\xi$-fluid dominated era leads to a substantial amplification of the primordial gravitational wave background, generating a prominent blue-tilted spectrum.
%During the non-standard expansion epoch (i.e., for modes re-entering the horizon at $a \leq a_{\text{rh}}$, corresponding to frequencies $f \geq f_{\text{rh}}$), the amplified GW spectrum, driven by the modified expansion history during the fluid-$\xi$ dominated era, vastly eclipses the standard inflationary contribution, generating a pronounced blue-tilted spectrum. 
Comparing the two panels of Fig.~\ref{fig:GW_spectrum}, it is evident that a larger EoS parameter ($w=0.7$ vs. $w=0.5$) produces a significantly steeper spectral tilt. Ultimately, this demonstrates that a post-inflationary reheating phase dominated by a stiff fluid amplifies the relic GW spectrum, drastically improving its detection prospects. Because our predicted amplified GW spectra peak predominantly in the low-to-mid frequency regime, both viable benchmark points fall comfortably within the sensitivity bands of future space-based observatories such as LISA, $\mu$-ARES, and THEIA.

\begin{figure}[tbp]
\centering 
\includegraphics[width=.49\textwidth]{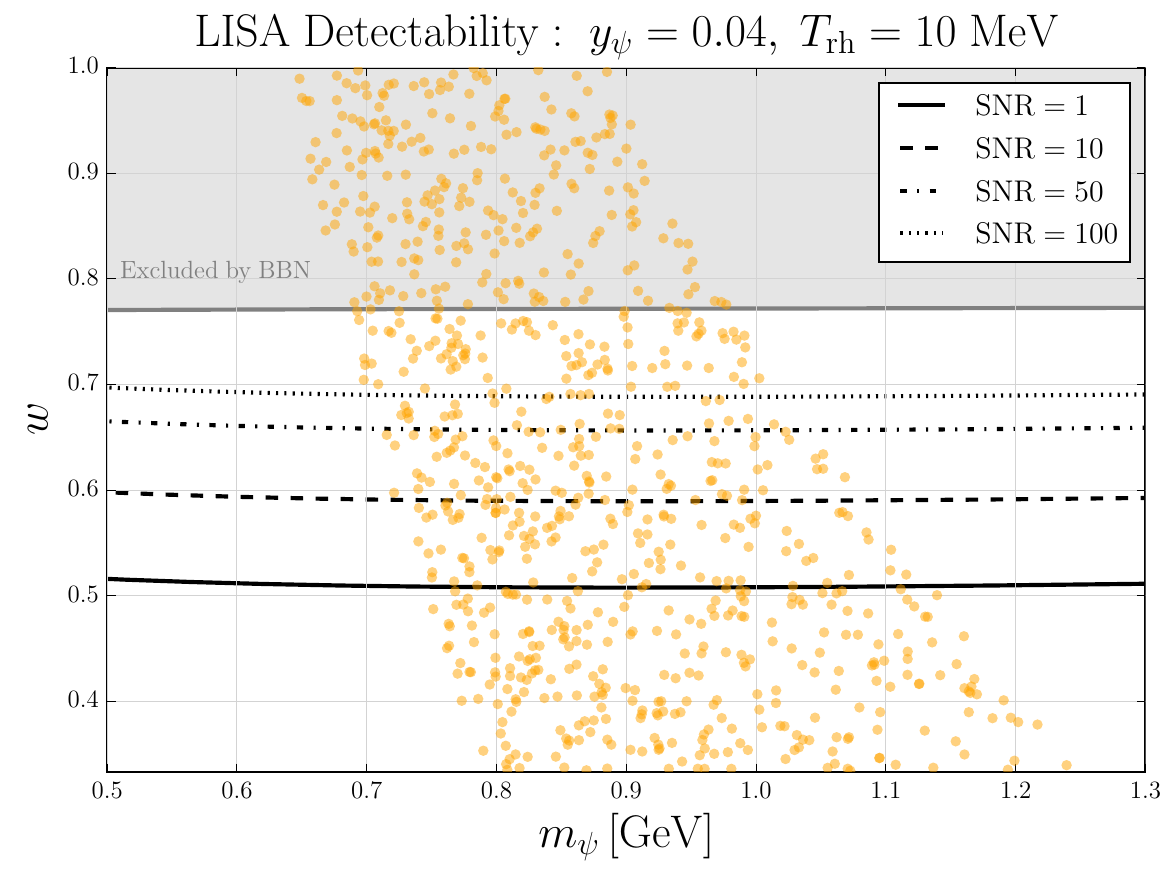}
\hfill
\includegraphics[width=.49\textwidth]{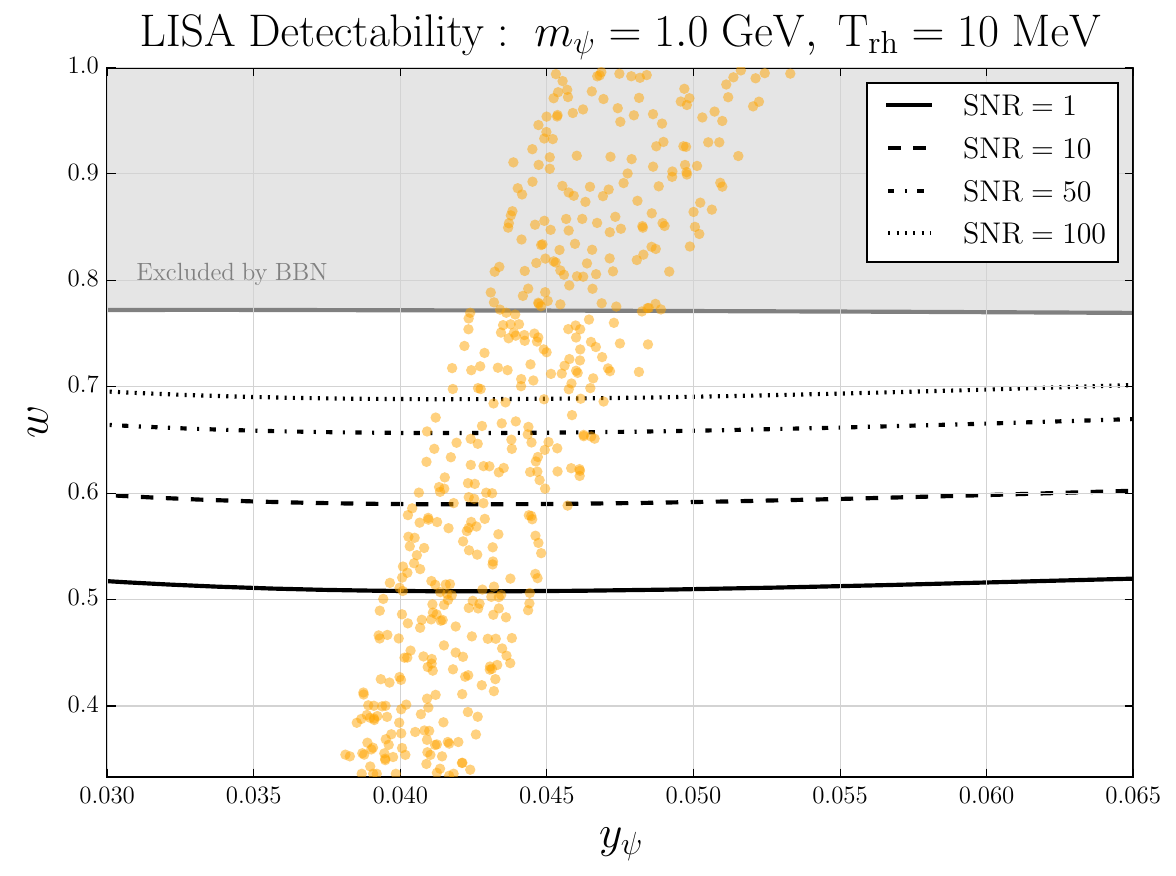}
\caption{Detectability of the viable parameter space at the upcoming LISA observatory. The displayed regions are simultaneously consistent with the observed light DM relic abundance, fall within the projected sensitivity of future terrestrial direct detection experiments, and strictly satisfy the cosmological $\Delta N_{\text{eff}}$ bounds. The left panel shows the $w-m_\psi$ plane for a fixed dark sector Yukawa coupling of $y_\psi=4 \times 10^{-2}$, while the right panel illustrates the $w-y_\psi$ plane for a fixed DM mass of $m_\psi = 1 \text{ GeV}$. The black contours indicate the expected Signal-to-Noise Ratio (SNR) for LISA~\cite{LISA:2017pwj}.} 
\label{fig:LISA_SNR}
\end{figure}

To identify the regions of parameter space that can be probed by future GW detectors, we compute the signal-to-noise ratio (SNR), defined as~\cite{Thrane:2013oya, Caprini:2015zlo}:\begin{equation}\text{SNR} = \sqrt{\tau_{\text{obs}} \int_{f_{\text{min}}}^{f_{\text{max}}} df \left(\frac{\Omega_{\text{GW}} h^2}{\Omega_{\text{GW}}^{\text{noise}} h^2}\right)^2}\,,\end{equation}where $\Omega_{\text{GW}}^{\text{noise}} h^2$ encodes the detector's effective noise spectrum over its sensitive frequency band $[f_{\text{min}}, f_{\text{max}}]$, and $\tau_{\text{obs}}$ is the total observation time. We evaluate the SNR for the parameter configurations that successfully reproduce the observed dark matter relic abundance (as shown in Fig.~\ref{fig:contourYDM}) in order to assess their GW detectability. Crucially, these highlighted regions of parameter space not only satisfy the proper thermal relic density requirements, but they also possess sufficiently large effective couplings to fall well within the projected sensitivities of upcoming terrestrial direct detection experiments. Furthermore, we strictly impose the current cosmological bounds on extra relativistic degrees of freedom, and any parameter combinations that generate excessive GW energy and violate the $\Delta N_{\text{eff}}$ constraint are excluded from the viable parameter space. As an illustrative example, Fig.~\ref{fig:LISA_SNR} displays the SNR contours for the LISA observatory~\cite{LISA:2017pwj}. The left panel illustrates SNR contours in the $m_\psi-w$ plane with a fixed Yukawa coupling of $y_\psi=4 \times 10^{-2}$, while the right panel shows the SNR contours in the $y_\psi-w$ plane for a fixed DM mass of $m_\psi = 1 \text{ GeV}$. The overlaid orange points lying in the unshaded regions indicate the scanned parameter space that is simultaneously consistent with the observed DM relic abundance, future direct detection reach, and the stringent $\Delta N_{\text{eff}}$ cosmological limits.

\section{Conclusion}\label{sec:conc}
The current challenges facing the traditional WIMP paradigm highlight the need for theoretical approaches that connect particle physics with early Universe cosmology. Sub-GeV LDM offers an exciting avenue for direct detection, but observable LDM with large effective couplings to SM often struggles with thermal underabundance and strict CMB limits on energy injection.

In this work, we explored how a minimal scalar portal model, utilizing a vector-like fermion dark matter candidate, can naturally ease the CMB tension due to its inherent p-wave annihilation suppression. Furthermore, to address the underabundant relic density typically caused by the large couplings needed for direct detection, we considered a pre-BBN non-standard cosmology driven by a stiff fluid ($w > 1/3$). The faster Hubble expansion during this epoch causes dark matter to freeze out earlier, which helps yield the correct relic abundance.

Importantly, this cosmological framework offers a potential observational signature. A stiff early epoch can blue-shift inflationary gravitational waves that re-enter the horizon before BBN, creating a characteristic high-frequency tilt in the stochastic gravitational wave background. This provides a fascinating correlation between the particle physics parameters targeted by terrestrial detectors, such as CRESST-III or SuperCDMS, and the signals that might be seen in future GW observatories like LISA and DECIGO. Ultimately, leveraging multi-messenger observations combining future direct detection results with gravitational wave data could offer us a much deeper understanding of both dark matter properties and the dynamics of pre-BBN cosmologies.

\acknowledgments
S.M. acknowledges support from the IIT Goa Startup Grant [2025/SG/SM/057]. K.Y.C. and E.L. acknowledge the financial support from National Research Foundation(NRF) grant
funded by the Korea government (MEST) NRF-2022R1A2C1005050.

\bibliographystyle{JCAP}
\bibliography{reference}

@book{Kolb:1990vq,
    author = "Kolb, Edward W. and Turner, Michael S.",
    title = "{The Early Universe}",
    reportNumber = "FERMILAB-BOOK-1990-01",
    doi = "10.1201/9780429492860",
    isbn = "978-0-429-49286-0, 978-0-201-62674-2",
    publisher = "Taylor and Francis",
    volume = "69",
    month = "5",
    year = "2019"
}

@article{Hoferichter:2017olk,
    author = "Hoferichter, Martin and Klos, Philipp and Men{\'e}ndez, Javier and Schwenk, Achim",
    title = "{Improved limits for Higgs-portal dark matter from LHC searches}",
    eprint = "1708.02245",
    archivePrefix = "arXiv",
    primaryClass = "hep-ph",
    reportNumber = "INT-PUB-17-031",
    doi = "10.1103/PhysRevLett.119.181803",
    journal = "Phys. Rev. Lett.",
    volume = "119",
    number = "18",
    pages = "181803",
    year = "2017"
}

@article{Bertone:2004pz,
    author = "Bertone, Gianfranco and Hooper, Dan and Silk, Joseph",
    title = "{Particle dark matter: Evidence, candidates and constraints}",
    eprint = "hep-ph/0404175",
    archivePrefix = "arXiv",
    reportNumber = "FERMILAB-PUB-04-047-A",
    doi = "10.1016/j.physrep.2004.08.031",
    journal = "Phys. Rept.",
    volume = "405",
    pages = "279--390",
    year = "2005"
}

@article{Ellis:2000ds,
    author = "Ellis, John R. and Ferstl, Andrew and Olive, Keith A.",
    title = "{Reevaluation of the elastic scattering of supersymmetric dark matter}",
    eprint = "hep-ph/0001005",
    archivePrefix = "arXiv",
    reportNumber = "CERN-TH-2000-001, UMN-TH-1834-2000, TPI-MINN-2000-01, CERN-TH-2K-001, TPI-MINN-00-01",
    doi = "10.1016/S0370-2693(00)00459-7",
    journal = "Phys. Lett. B",
    volume = "481",
    pages = "304--314",
    year = "2000"
}

@article{ParticleDataGroup:2020ssz,
    author = "Zyla, P. A. and others",
    collaboration = "Particle Data Group",
    title = "{Review of Particle Physics}",
    doi = "10.1093/ptep/ptaa104",
    journal = "PTEP",
    volume = "2020",
    number = "8",
    pages = "083C01",
    year = "2020"
}

@article{PandaX:2022xqx,
    author = "Li, Shuaijie and others",
    collaboration = "PandaX",
    title = "{Search for Light Dark Matter with Ionization Signals in the PandaX-4T Experiment}",
    eprint = "2212.10067",
    archivePrefix = "arXiv",
    primaryClass = "hep-ex",
    doi = "10.1103/PhysRevLett.130.261001",
    journal = "Phys. Rev. Lett.",
    volume = "130",
    number = "26",
    pages = "261001",
    year = "2023"
}

@article{XENON:2019gfn,
    author = "Aprile, E. and others",
    collaboration = "XENON",
    title = "{Light Dark Matter Search with Ionization Signals in XENON1T}",
    eprint = "1907.11485",
    archivePrefix = "arXiv",
    primaryClass = "hep-ex",
    doi = "10.1103/PhysRevLett.123.251801",
    journal = "Phys. Rev. Lett.",
    volume = "123",
    number = "25",
    pages = "251801",
    year = "2019"
}

@article{GlobalArgonDarkMatter:2022ppc,
    author = "Agnes, P. and others",
    collaboration = "Global Argon Dark Matter",
    title = "{Sensitivity projections for a dual-phase argon TPC optimized for light dark matter searches through the ionization channel}",
    eprint = "2209.01177",
    archivePrefix = "arXiv",
    primaryClass = "physics.ins-det",
    doi = "10.1103/PhysRevD.107.112006",
    journal = "Phys. Rev. D",
    volume = "107",
    number = "11",
    pages = "112006",
    year = "2023"
}

@article{DarkSide-50:2022qzh,
    author = "Agnes, P. and others",
    collaboration = "DarkSide-50",
    title = "{Search for low-mass dark matter WIMPs with 12~ton-day exposure of DarkSide-50}",
    eprint = "2207.11966",
    archivePrefix = "arXiv",
    primaryClass = "hep-ex",
    reportNumber = "FERMILAB-PUB-22-589-ND-PPD-SCD",
    doi = "10.1103/PhysRevD.107.063001",
    journal = "Phys. Rev. D",
    volume = "107",
    number = "6",
    pages = "063001",
    year = "2023"
}

@article{DAMIC-M:2025luv,
    author = "Aggarwal, K. and others",
    collaboration = "DAMIC-M",
    title = "{Probing Benchmark Models of Hidden-Sector Dark Matter with DAMIC-M}",
    eprint = "2503.14617",
    archivePrefix = "arXiv",
    primaryClass = "hep-ex",
    month = "3",
    year = "2025"
}

@article{SuperCDMS:2024yiv,
    author = "Albakry, M. F. and others",
    collaboration = "SuperCDMS",
    title = "{Light dark matter constraints from SuperCDMS HVeV detectors operated underground with an anticoincidence event selection}",
    eprint = "2407.08085",
    archivePrefix = "arXiv",
    primaryClass = "hep-ex",
    reportNumber = "FERMILAB-PUB-24-0376-PPD",
    doi = "10.1103/PhysRevD.111.012006",
    journal = "Phys. Rev. D",
    volume = "111",
    number = "1",
    pages = "012006",
    year = "2025"
}

@article{PandaX-II:2021nsg,
    author = "Cheng, Chen and others",
    collaboration = "PandaX-II",
    title = "{Search for Light Dark Matter-Electron Scatterings in the PandaX-II Experiment}",
    eprint = "2101.07479",
    archivePrefix = "arXiv",
    primaryClass = "hep-ex",
    doi = "10.1103/PhysRevLett.126.211803",
    journal = "Phys. Rev. Lett.",
    volume = "126",
    number = "21",
    pages = "211803",
    year = "2021"
}

@article{CRESST:2019jnq,
    author = "Abdelhameed, A. H. and others",
    collaboration = "CRESST",
    title = "{First results from the CRESST-III low-mass dark matter program}",
    eprint = "1904.00498",
    archivePrefix = "arXiv",
    primaryClass = "astro-ph.CO",
    doi = "10.1103/PhysRevD.100.102002",
    journal = "Phys. Rev. D",
    volume = "100",
    number = "10",
    pages = "102002",
    year = "2019"
}

@article{SENSEI:2020dpa,
    author = "Barak, Liron and others",
    collaboration = "SENSEI",
    title = "{SENSEI: Direct-Detection Results on sub-GeV Dark Matter from a New Skipper-CCD}",
    eprint = "2004.11378",
    archivePrefix = "arXiv",
    primaryClass = "astro-ph.CO",
    reportNumber = "YITP-SB-2020-6, FERMILAB-PUB-20-158-AE-E",
    doi = "10.1103/PhysRevLett.125.171802",
    journal = "Phys. Rev. Lett.",
    volume = "125",
    number = "17",
    pages = "171802",
    year = "2020"
}

@article{Baer:2014eja,
    author = "Baer, Howard and Choi, Ki-Young and Kim, Jihn E. and Roszkowski, Leszek",
    title = "{Dark matter production in the early Universe: beyond the thermal WIMP paradigm}",
    eprint = "1407.0017",
    archivePrefix = "arXiv",
    primaryClass = "hep-ph",
    doi = "10.1016/j.physrep.2014.10.002",
    journal = "Phys. Rept.",
    volume = "555",
    pages = "1--60",
    year = "2015"
}

@article{COSINE-100:2021poy,
    author = "Adhikari, G. and others",
    collaboration = "COSINE-100",
    title = "{Searching for low-mass dark matter via the Migdal effect in COSINE-100}",
    eprint = "2110.05806",
    archivePrefix = "arXiv",
    primaryClass = "hep-ex",
    doi = "10.1103/PhysRevD.105.042006",
    journal = "Phys. Rev. D",
    volume = "105",
    number = "4",
    pages = "042006",
    year = "2022"
}

@article{XENON:2024znc,
    author = "Aprile, E. and others",
    collaboration = "XENON",
    title = "{Search for Light Dark Matter in Low-Energy Ionization Signals from XENONnT}",
    eprint = "2411.15289",
    archivePrefix = "arXiv",
    primaryClass = "hep-ex",
    doi = "10.1103/PhysRevLett.134.161004",
    journal = "Phys. Rev. Lett.",
    volume = "134",
    number = "16",
    pages = "161004",
    year = "2025"
}

@article{Balan:2024cmq,
    author = "Balan, Sowmiya and others",
    title = "{Resonant or asymmetric: the status of sub-GeV dark matter}",
    eprint = "2405.17548",
    archivePrefix = "arXiv",
    primaryClass = "hep-ph",
    reportNumber = "TTP24-015, P3H-24-033",
    doi = "10.1088/1475-7516/2025/01/053",
    journal = "JCAP",
    volume = "01",
    pages = "053",
    year = "2025"
}

@article{Cheek:2025nul,
    author = "Cheek, Andrew and Figueroa, Pablo and Herrera, Gonzalo and Shoemaker, Ian M.",
    title = "{Sub-GeV Dark Matter Under Pressure from Direct Detection}",
    eprint = "2507.15956",
    archivePrefix = "arXiv",
    primaryClass = "hep-ph",
    month = "7",
    year = "2025"
}

@article{Borah:2024yow,
    author = "Borah, Debasish and Das, Pritam and Mahapatra, Satyabrata and Sahu, Narendra",
    title = "{Light thermal dark matter via type-I seesaw portal}",
    eprint = "2401.01639",
    archivePrefix = "arXiv",
    primaryClass = "hep-ph",
    doi = "10.1007/JHEP08(2025)023",
    journal = "JHEP",
    volume = "08",
    pages = "023",
    year = "2025"
}

@article{Krnjaic:2025noj,
    author = "Krnjaic, Gordan",
    title = "{Testing Thermal-Relic Dark Matter with a Dark Photon Mediator}",
    eprint = "2505.04626",
    archivePrefix = "arXiv",
    primaryClass = "hep-ph",
    reportNumber = "FERMILAB-PUB-25-0234-T",
    month = "5",
    year = "2025"
}

@article{Adhikary:2024btd,
    author = "Adhikary, Amit and Borah, Debasish and Mahapatra, Satyabrata and Saha, Indrajit and Sahu, Narendra and Thounaojam, Vicky Singh",
    title = "{New realisation of light thermal dark matter with enhanced detection prospects}",
    eprint = "2405.17564",
    archivePrefix = "arXiv",
    primaryClass = "hep-ph",
    doi = "10.1088/1475-7516/2024/12/043",
    journal = "JCAP",
    volume = "12",
    pages = "043",
    year = "2024"
}

@article{Dutta:2019fxn,
    author = "Dutta, Bhaskar and Ghosh, Sumit and Kumar, Jason",
    title = "{A sub-GeV dark matter model}",
    eprint = "1905.02692",
    archivePrefix = "arXiv",
    primaryClass = "hep-ph",
    reportNumber = "MI-TH-1921, UH511-1305-2019",
    doi = "10.1103/PhysRevD.100.075028",
    journal = "Phys. Rev. D",
    volume = "100",
    pages = "075028",
    year = "2019"
}

@article{Essig:2017kqs,
    author = "Essig, Rouven and Volansky, Tomer and Yu, Tien-Tien",
    title = "{New Constraints and Prospects for sub-GeV Dark Matter Scattering off Electrons in Xenon}",
    eprint = "1703.00910",
    archivePrefix = "arXiv",
    primaryClass = "hep-ph",
    reportNumber = "CERN-TH-2017-042, YITP-SB-17-09",
    doi = "10.1103/PhysRevD.96.043017",
    journal = "Phys. Rev. D",
    volume = "96",
    number = "4",
    pages = "043017",
    year = "2017"
}

@article{Bondarenko:2019vrb,
    author = "Bondarenko, Kyrylo and Boyarsky, Alexey and Bringmann, Torsten and Hufnagel, Marco and Schmidt-Hoberg, Kai and Sokolenko, Anastasia",
    title = "{Direct detection and complementary constraints for sub-GeV dark matter}",
    eprint = "1909.08632",
    archivePrefix = "arXiv",
    primaryClass = "hep-ph",
    reportNumber = "DESY-19-140",
    doi = "10.1007/JHEP03(2020)118",
    journal = "JHEP",
    volume = "03",
    pages = "118",
    year = "2020"
}

@article{Borah:2025wcc,
    author = "Borah, Debasish and Mahapatra, Satyabrata and Sahu, Narendra and Thounaojam, Vicky Singh",
    title = "{Light thermal dark matter models in the light of DAMIC-M 2025 constraints}",
    eprint = "2509.16319",
    archivePrefix = "arXiv",
    primaryClass = "hep-ph",
    doi = "10.1103/zv1q-3j7c",
    journal = "Phys. Rev. D",
    volume = "113",
    number = "1",
    pages = "015026",
    year = "2026"
}

@article{Elor:2021swj,
    author = "Elor, Gilly and McGehee, Robert and Pierce, Aaron",
    title = "{Maximizing Direct Detection with Highly Interactive Particle Relic Dark Matter}",
    eprint = "2112.03920",
    archivePrefix = "arXiv",
    primaryClass = "hep-ph",
    reportNumber = "LCTP-21-26, MITP-21-044",
    doi = "10.1103/PhysRevLett.130.031803",
    journal = "Phys. Rev. Lett.",
    volume = "130",
    number = "3",
    pages = "031803",
    year = "2023"
}

@article{Schumann:2019eaa,
    author = "Schumann, Marc",
    title = "{Direct Detection of WIMP Dark Matter: Concepts and Status}",
    eprint = "1903.03026",
    archivePrefix = "arXiv",
    primaryClass = "astro-ph.CO",
    doi = "10.1088/1361-6471/ab2ea5",
    journal = "J. Phys. G",
    volume = "46",
    number = "10",
    pages = "103003",
    year = "2019"
}

@article{Zurek:2024qfm,
    author = "Zurek, Kathryn M.",
    title = "{Dark Matter Candidates of a Very Low Mass}",
    eprint = "2401.03025",
    archivePrefix = "arXiv",
    primaryClass = "hep-ph",
    doi = "10.1146/annurev-nucl-101918-023542",
    journal = "Ann. Rev. Nucl. Part. Sci.",
    volume = "74",
    number = "1",
    pages = "287--319",
    year = "2024"
}

@article{Fermi-LAT:2015att,
    author = "Ackermann, M. and others",
    collaboration = "Fermi-LAT",
    title = "{Searching for Dark Matter Annihilation from Milky Way Dwarf Spheroidal Galaxies with Six Years of Fermi Large Area Telescope Data}",
    eprint = "1503.02641",
    archivePrefix = "arXiv",
    primaryClass = "astro-ph.HE",
    reportNumber = "FERMILAB-PUB-15-081-AE",
    doi = "10.1103/PhysRevLett.115.231301",
    journal = "Phys. Rev. Lett.",
    volume = "115",
    number = "23",
    pages = "231301",
    year = "2015"
}

@article{HESS:2018cbt,
    author = "Abdallah, H. and others",
    collaboration = "HESS",
    title = "{Search for $\gamma$-Ray Line Signals from Dark Matter Annihilations in the Inner Galactic Halo from 10 Years of Observations with H.E.S.S.}",
    eprint = "1805.05741",
    archivePrefix = "arXiv",
    primaryClass = "astro-ph.HE",
    doi = "10.1103/PhysRevLett.120.201101",
    journal = "Phys. Rev. Lett.",
    volume = "120",
    number = "20",
    pages = "201101",
    year = "2018"
}

@article{Profumo:2017obk,
    author = "Profumo, Stefano and Queiroz, Farinaldo S. and Silk, Joseph and Siqueira, Clarissa",
    title = "{Searching for Secluded Dark Matter with H.E.S.S., Fermi-LAT, and Planck}",
    eprint = "1711.03133",
    archivePrefix = "arXiv",
    primaryClass = "hep-ph",
    doi = "10.1088/1475-7516/2018/03/010",
    journal = "JCAP",
    volume = "03",
    pages = "010",
    year = "2018"
}

@article{Madhavacheril:2013cna,
    author = "Madhavacheril, Mathew S. and Sehgal, Neelima and Slatyer, Tracy R.",
    title = "{Current Dark Matter Annihilation Constraints from CMB and Low-Redshift Data}",
    eprint = "1310.3815",
    archivePrefix = "arXiv",
    primaryClass = "astro-ph.CO",
    reportNumber = "MIT-CTP-4505",
    doi = "10.1103/PhysRevD.89.103508",
    journal = "Phys. Rev. D",
    volume = "89",
    pages = "103508",
    year = "2014"
}

@article{Slatyer:2015jla,
    author = "Slatyer, Tracy R.",
    title = "{Indirect dark matter signatures in the cosmic dark ages. I. Generalizing the bound on s-wave dark matter annihilation from Planck results}",
    eprint = "1506.03811",
    archivePrefix = "arXiv",
    primaryClass = "hep-ph",
    reportNumber = "MIT-CTP-4682",
    doi = "10.1103/PhysRevD.93.023527",
    journal = "Phys. Rev. D",
    volume = "93",
    number = "2",
    pages = "023527",
    year = "2016"
}

@article{Elor:2015bho,
    author = "Elor, Gilly and Rodd, Nicholas L. and Slatyer, Tracy R. and Xue, Wei",
    title = "{Model-Independent Indirect Detection Constraints on Hidden Sector Dark Matter}",
    eprint = "1511.08787",
    archivePrefix = "arXiv",
    primaryClass = "hep-ph",
    reportNumber = "MIT-CTP-4742",
    doi = "10.1088/1475-7516/2016/06/024",
    journal = "JCAP",
    volume = "06",
    pages = "024",
    year = "2016"
}

@article{Arcadi:2024jzv,
    author = "Arcadi, Giorgio",
    title = "{Thermal and non-thermal DM production in non-Standard Cosmologies: a mini review}",
    eprint = "2406.11042",
    archivePrefix = "arXiv",
    primaryClass = "hep-ph",
    month = "6",
    year = "2024"
}

@article{Arias:2019uol,
    author = "Arias, Paola and Bernal, Nicol{\'a}s and Herrera, Alan and Maldonado, Carlos",
    title = "{Reconstructing Non-standard Cosmologies with Dark Matter}",
    eprint = "1906.04183",
    archivePrefix = "arXiv",
    primaryClass = "hep-ph",
    reportNumber = "PI/UAN-2019-649FT",
    doi = "10.1088/1475-7516/2019/10/047",
    journal = "JCAP",
    volume = "10",
    pages = "047",
    year = "2019"
}

@article{Soman:2024zor,
    author = "Soman, Athul K. and Mishra, Swagat S. and Shafi, Mohammed and Basak, Soumen",
    title = "{Inflationary gravitational waves as a probe of the unknown postinflationary primordial Universe}",
    eprint = "2407.07956",
    archivePrefix = "arXiv",
    primaryClass = "gr-qc",
    doi = "10.1103/81wm-jlm6",
    journal = "Phys. Rev. D",
    volume = "112",
    number = "10",
    pages = "103521",
    year = "2025"
}

@article{Mishra:2025nnu,
    author = "Mishra, Swagat S. and Soman, Athul K.",
    title = "{Morphology of Inflationary Gravitational Wave Spectra imprinted by a Sequence of Post-Inflationary Epochs $via$${\rm GWInSpect}$}",
    eprint = "2510.25672",
    archivePrefix = "arXiv",
    primaryClass = "astro-ph.CO",
    month = "10",
    year = "2025"
}

@article{Konings:2024zvz,
    author = "Konings, Annet and Marinichenko, Mariia and Mikulenko, Oleksii and Patil, Subodh P.",
    title = "{Primordial Gravitational Wave Probes of Non-Standard Thermal Histories}",
    eprint = "2412.15144",
    archivePrefix = "arXiv",
    primaryClass = "astro-ph.CO",
    month = "12",
    year = "2024"
}

@article{ATLAS:2022yvh,
    author = "Aad, Georges and others",
    collaboration = "ATLAS",
    title = "{Search for invisible Higgs-boson decays in events with vector-boson fusion signatures using 139 fb$^{-1}$ of proton-proton data recorded by the ATLAS experiment}",
    eprint = "2202.07953",
    archivePrefix = "arXiv",
    primaryClass = "hep-ex",
    reportNumber = "CERN-EP-2021-258",
    doi = "10.1007/JHEP08(2022)104",
    journal = "JHEP",
    volume = "08",
    pages = "104",
    year = "2022"
}

@article{Balaji:2022noj,
    author = "Balaji, Shyam and Dev, P. S. Bhupal and Silk, Joseph and Zhang, Yongchao",
    title = "{Improved stellar limits on a light CP-even scalar}",
    eprint = "2205.01669",
    archivePrefix = "arXiv",
    primaryClass = "hep-ph",
    doi = "10.1088/1475-7516/2022/12/024",
    journal = "JCAP",
    volume = "12",
    pages = "024",
    year = "2022"
}

@article{Dev:2017dui,
    author = "Dev, P. S. Bhupal and Mohapatra, Rabindra N. and Zhang, Yongchao",
    title = "{Long Lived Light Scalars as Probe of Low Scale Seesaw Models}",
    eprint = "1703.02471",
    archivePrefix = "arXiv",
    primaryClass = "hep-ph",
    reportNumber = "ULB-TH-17-05, UMD-PP-017-21",
    doi = "10.1016/j.nuclphysb.2017.07.021",
    journal = "Nucl. Phys. B",
    volume = "923",
    pages = "179--221",
    year = "2017"
}

@article{Dev:2019hho,
    author = "Dev, P. S. Bhupal and Mohapatra, Rabindra N. and Zhang, Yongchao",
    title = "{Constraints on long-lived light scalars with flavor-changing couplings and the KOTO anomaly}",
    eprint = "1911.12334",
    archivePrefix = "arXiv",
    primaryClass = "hep-ph",
    doi = "10.1103/PhysRevD.101.075014",
    journal = "Phys. Rev. D",
    volume = "101",
    number = "7",
    pages = "075014",
    year = "2020"
}

@article{Egana-Ugrinovic:2019wzj,
    author = "Egana-Ugrinovic, Daniel and Homiller, Samuel and Meade, Patrick",
    title = "{Light Scalars and the Koto Anomaly}",
    eprint = "1911.10203",
    archivePrefix = "arXiv",
    primaryClass = "hep-ph",
    reportNumber = "YITP-SB-19-43",
    doi = "10.1103/PhysRevLett.124.191801",
    journal = "Phys. Rev. Lett.",
    volume = "124",
    number = "19",
    pages = "191801",
    year = "2020"
}

@article{Batell:2019nwo,
    author = "Batell, Brian and Berger, Joshua and Ismail, Ahmed",
    title = "{Probing the Higgs Portal at the Fermilab Short-Baseline Neutrino Experiments}",
    eprint = "1909.11670",
    archivePrefix = "arXiv",
    primaryClass = "hep-ph",
    reportNumber = "PITT-PACC-1906, OSU-HEP-19-07",
    doi = "10.1103/PhysRevD.100.115039",
    journal = "Phys. Rev. D",
    volume = "100",
    number = "11",
    pages = "115039",
    year = "2019"
}

@article{Berryman:2019dme,
    author = "Berryman, Jeffrey M. and de Gouvea, Andre and Fox, Patrick J and Kayser, Boris Jules and Kelly, Kevin James and Raaf, Jennifer Lynne",
    title = "{Searches for Decays of New Particles in the DUNE Multi-Purpose Near Detector}",
    eprint = "1912.07622",
    archivePrefix = "arXiv",
    primaryClass = "hep-ph",
    reportNumber = "FERMILAB-PUB-19-607-ND-T, NUHEP-TH/19-16",
    doi = "10.1007/JHEP02(2020)174",
    journal = "JHEP",
    volume = "02",
    pages = "174",
    year = "2020"
}

@article{Anchordoqui:2021ghd,
    author = "Anchordoqui, Luis A. and others",
    title = "{The Forward Physics Facility: Sites, experiments, and physics potential}",
    eprint = "2109.10905",
    archivePrefix = "arXiv",
    primaryClass = "hep-ph",
    reportNumber = "BNL-222142-2021-FORE, CERN-PBC-Notes-2021-025, DESY-21-142, DESY-21-142,
  FERMILAB-CONF-21-452-AE-E-ND-PPD-T, KYUSHU-RCAPP-2021-01, LU TP 21-36,
  PITT-PACC-2118, SMU-HEP-21-10, UCI-TR-2021-22, FERMILAB-CONF-21-452-AE-E-ND-PPD-T",
    doi = "10.1016/j.physrep.2022.04.004",
    journal = "Phys. Rept.",
    volume = "968",
    pages = "1--50",
    year = "2022"
}

@article{Watanabe:2006qe,
    author = "Watanabe, Yuki and Komatsu, Eiichiro",
    title = "{Improved Calculation of the Primordial Gravitational Wave Spectrum in the Standard Model}",
    eprint = "astro-ph/0604176",
    archivePrefix = "arXiv",
    doi = "10.1103/PhysRevD.73.123515",
    journal = "Phys. Rev. D",
    volume = "73",
    pages = "123515",
    year = "2006"
}

@article{Hannestad:2004px,
    author = "Hannestad, Steen",
    title = "{What is the lowest possible reheating temperature?}",
    eprint = "astro-ph/0403291",
    archivePrefix = "arXiv",
    doi = "10.1103/PhysRevD.70.043506",
    journal = "Phys. Rev. D",
    volume = "70",
    pages = "043506",
    year = "2004"
}

@article{Planck:2018vyg,
    author = "Aghanim, N. and others",
    collaboration = "Planck",
    title = "{Planck 2018 results. VI. Cosmological parameters}",
    eprint = "1807.06209",
    archivePrefix = "arXiv",
    primaryClass = "astro-ph.CO",
    doi = "10.1051/0004-6361/201833910",
    journal = "Astron. Astrophys.",
    volume = "641",
    pages = "A6",
    year = "2020",
    note = "[Erratum: Astron.Astrophys. 652, C4 (2021)]"
}

@article{Planck:2018jri,
    author = "Akrami, Y. and others",
    collaboration = "Planck",
    title = "{Planck 2018 results. X. Constraints on inflation}",
    eprint = "1807.06211",
    archivePrefix = "arXiv",
    primaryClass = "astro-ph.CO",
    doi = "10.1051/0004-6361/201833887",
    journal = "Astron. Astrophys.",
    volume = "641",
    pages = "A10",
    year = "2020"
}

@article{CMB-S4:2016ple,
    author = "Abazajian, Kevork N. and others",
    collaboration = "CMB-S4",
    title = "{CMB-S4 Science Book, First Edition}",
    eprint = "1610.02743",
    archivePrefix = "arXiv",
    primaryClass = "astro-ph.CO",
    reportNumber = "FERMILAB-FN-1024-A-AE",
    month = "10",
    year = "2016"
}

@article{CMB-HD:2022bsz,
    author = "Aiola, Simone and others",
    collaboration = "CMB-HD",
    title = "{Snowmass2021 CMB-HD White Paper}",
    eprint = "2203.05728",
    archivePrefix = "arXiv",
    primaryClass = "astro-ph.CO",
    reportNumber = "FERMILAB-PUB-22-344-PPD",
    month = "3",
    year = "2022"
}

@article{Mangano:2005cc,
    author = "Mangano, Gianpiero and Miele, Gennaro and Pastor, Sergio and Pinto, Teguayco and Pisanti, Ofelia and Serpico, Pasquale D.",
    title = "{Relic neutrino decoupling including flavor oscillations}",
    eprint = "hep-ph/0506164",
    archivePrefix = "arXiv",
    reportNumber = "DSF-16-2005, IFIC-05-17, MPP-2005-36",
    doi = "10.1016/j.nuclphysb.2005.09.041",
    journal = "Nucl. Phys. B",
    volume = "729",
    pages = "221--234",
    year = "2005"
}

@article{Reitze:2019iox,
    author = "Reitze, David and others",
    title = "{Cosmic Explorer: The U.S. Contribution to Gravitational-Wave Astronomy beyond LIGO}",
    eprint = "1907.04833",
    archivePrefix = "arXiv",
    primaryClass = "astro-ph.IM",
    reportNumber = "LIGO-P1900316",
    journal = "Bull. Am. Astron. Soc.",
    volume = "51",
    number = "7",
    pages = "035",
    year = "2019"
}

@article{Turner:1993vb,
    author = "Turner, Michael S. and White, Martin J. and Lidsey, James E.",
    title = "{Tensor perturbations in inflationary models as a probe of cosmology}",
    eprint = "astro-ph/9306029",
    archivePrefix = "arXiv",
    reportNumber = "FERMILAB-PUB-93-069-A, CFPA-TH-93-19, CFPA-93-19",
    doi = "10.1103/PhysRevD.48.4613",
    journal = "Phys. Rev. D",
    volume = "48",
    pages = "4613--4622",
    year = "1993"
}

@article{Chongchitnan:2006pe,
    author = "Chongchitnan, Sirichai and Efstathiou, George",
    title = "{Prospects for direct detection of primordial gravitational waves}",
    eprint = "astro-ph/0602594",
    archivePrefix = "arXiv",
    doi = "10.1103/PhysRevD.73.083511",
    journal = "Phys. Rev. D",
    volume = "73",
    pages = "083511",
    year = "2006"
}

@article{Nakayama:2008wy,
    author = "Nakayama, Kazunori and Saito, Shun and Suwa, Yudai and Yokoyama, Jun'ichi",
    title = "{Probing reheating temperature of the universe with gravitational wave background}",
    eprint = "0804.1827",
    archivePrefix = "arXiv",
    primaryClass = "astro-ph",
    reportNumber = "RESCEU-7-08, UTAP-596",
    doi = "10.1088/1475-7516/2008/06/020",
    journal = "JCAP",
    volume = "06",
    pages = "020",
    year = "2008"
}

@article{Kuroyanagi:2011fy,
    author = "Kuroyanagi, Sachiko and Nakayama, Kazunori and Saito, Shun",
    title = "{Prospects for determination of thermal history after inflation with future gravitational wave detectors}",
    eprint = "1110.4169",
    archivePrefix = "arXiv",
    primaryClass = "astro-ph.CO",
    reportNumber = "ICRR-REPORT-597-2011-14, UT-11-35",
    doi = "10.1103/PhysRevD.84.123513",
    journal = "Phys. Rev. D",
    volume = "84",
    pages = "123513",
    year = "2011"
}

@article{Kuroyanagi:2014nba,
    author = "Kuroyanagi, Sachiko and Takahashi, Tomo and Yokoyama, Shuichiro",
    title = "{Blue-tilted Tensor Spectrum and Thermal History of the Universe}",
    eprint = "1407.4785",
    archivePrefix = "arXiv",
    primaryClass = "astro-ph.CO",
    reportNumber = "ICRR-REPORT-686-2014-12",
    doi = "10.1088/1475-7516/2015/02/003",
    journal = "JCAP",
    volume = "02",
    pages = "003",
    year = "2015"
}

@article{Ghoshal:2025ldb,
    author = "Ghoshal, Anish and Paul, Debarun and Pal, Supratik",
    title = "{Primordial Gravitational Waves as Complementary Probe of Dark Matter Indirect Detection}",
    eprint = "2506.17568",
    archivePrefix = "arXiv",
    primaryClass = "hep-ph",
    month = "6",
    year = "2025"
}

@article{Caprini:2018mtu,
    author = "Caprini, Chiara and Figueroa, Daniel G.",
    title = "{Cosmological Backgrounds of Gravitational Waves}",
    eprint = "1801.04268",
    archivePrefix = "arXiv",
    primaryClass = "astro-ph.CO",
    doi = "10.1088/1361-6382/aac608",
    journal = "Class. Quant. Grav.",
    volume = "35",
    number = "16",
    pages = "163001",
    year = "2018"
}

@article{Boyle:2005se,
    author = "Boyle, Latham A. and Steinhardt, Paul J.",
    title = "{Probing the early universe with inflationary gravitational waves}",
    eprint = "astro-ph/0512014",
    archivePrefix = "arXiv",
    doi = "10.1103/PhysRevD.77.063504",
    journal = "Phys. Rev. D",
    volume = "77",
    pages = "063504",
    year = "2008"
}

@article{Crowder:2005nr,
    author = "Crowder, Jeff and Cornish, Neil J.",
    title = "{Beyond LISA: Exploring future gravitational wave missions}",
    eprint = "gr-qc/0506015",
    archivePrefix = "arXiv",
    doi = "10.1103/PhysRevD.72.083005",
    journal = "Phys. Rev. D",
    volume = "72",
    pages = "083005",
    year = "2005"
}

@article{Corbin:2005ny,
    author = "Corbin, Vincent and Cornish, Neil J.",
    title = "{Detecting the cosmic gravitational wave background with the big bang observer}",
    eprint = "gr-qc/0512039",
    archivePrefix = "arXiv",
    doi = "10.1088/0264-9381/23/7/014",
    journal = "Class. Quant. Grav.",
    volume = "23",
    pages = "2435--2446",
    year = "2006"
}

@article{Seto:2001qf,
    author = "Seto, Naoki and Kawamura, Seiji and Nakamura, Takashi",
    title = "{Possibility of direct measurement of the acceleration of the universe using 0.1-Hz band laser interferometer gravitational wave antenna in space}",
    eprint = "astro-ph/0108011",
    archivePrefix = "arXiv",
    doi = "10.1103/PhysRevLett.87.221103",
    journal = "Phys. Rev. Lett.",
    volume = "87",
    pages = "221103",
    year = "2001"
}

@article{Kudoh:2005as,
    author = "Kudoh, Hideaki and Taruya, Atsushi and Hiramatsu, Takashi and Himemoto, Yoshiaki",
    title = "{Detecting a gravitational-wave background with next-generation space interferometers}",
    eprint = "gr-qc/0511145",
    archivePrefix = "arXiv",
    reportNumber = "UTAP-544, RESCEU-37-05",
    doi = "10.1103/PhysRevD.73.064006",
    journal = "Phys. Rev. D",
    volume = "73",
    pages = "064006",
    year = "2006"
}

@article{LISA:2017pwj,
    author = "Amaro-Seoane, Pau and others",
    collaboration = "LISA",
    title = "{Laser Interferometer Space Antenna}",
    eprint = "1702.00786",
    archivePrefix = "arXiv",
    primaryClass = "astro-ph.IM",
    month = "2",
    year = "2017"
}

@article{Sesana:2019vho,
    author = "Sesana, Alberto and others",
    title = "{Unveiling the gravitational universe at $\mu$-Hz frequencies}",
    eprint = "1908.11391",
    archivePrefix = "arXiv",
    primaryClass = "astro-ph.IM",
    doi = "10.1007/s10686-021-09709-9",
    journal = "Exper. Astron.",
    volume = "51",
    number = "3",
    pages = "1333--1383",
    year = "2021"
}

@article{Sathyaprakash:2012jk,
    author = "Sathyaprakash, B. and others",
    editor = "Hannam, Mark and Sutton, Patrick and Hild, Stefan and van den Broeck, Chris",
    title = "{Scientific Objectives of Einstein Telescope}",
    eprint = "1206.0331",
    archivePrefix = "arXiv",
    primaryClass = "gr-qc",
    doi = "10.1088/0264-9381/29/12/124013",
    journal = "Class. Quant. Grav.",
    volume = "29",
    pages = "124013",
    year = "2012",
    note = "[Erratum: Class.Quant.Grav. 30, 079501 (2013)]"
}

@article{ET:2019dnz,
    author = "Maggiore, Michele and others",
    collaboration = "ET",
    title = "{Science Case for the Einstein Telescope}",
    eprint = "1912.02622",
    archivePrefix = "arXiv",
    primaryClass = "astro-ph.CO",
    doi = "10.1088/1475-7516/2020/03/050",
    journal = "JCAP",
    volume = "03",
    pages = "050",
    year = "2020"
}

@article{Garcia-Bellido:2021zgu,
    author = "Garcia-Bellido, Juan and Murayama, Hitoshi and White, Graham",
    title = "{Exploring the early Universe with Gaia and Theia}",
    eprint = "2104.04778",
    archivePrefix = "arXiv",
    primaryClass = "hep-ph",
    reportNumber = "IFT-UAM/CSIC-2021-038",
    doi = "10.1088/1475-7516/2021/12/023",
    journal = "JCAP",
    volume = "12",
    number = "12",
    pages = "023",
    year = "2021"
}

@article{LIGOScientific:2014pky,
    author = "Aasi, J. and others",
    collaboration = "LIGO Scientific",
    title = "{Advanced LIGO}",
    eprint = "1411.4547",
    archivePrefix = "arXiv",
    primaryClass = "gr-qc",
    doi = "10.1088/0264-9381/32/7/074001",
    journal = "Class. Quant. Grav.",
    volume = "32",
    pages = "074001",
    year = "2015"
}

@article{CDEX:2019hzn,
    author = "Liu, Z. Z. and others",
    collaboration = "CDEX",
    title = "{Constraints on Spin-Independent Nucleus Scattering with sub-GeV Weakly Interacting Massive Particle Dark Matter from the CDEX-1B Experiment at the China Jinping Underground Laboratory}",
    eprint = "1905.00354",
    archivePrefix = "arXiv",
    primaryClass = "hep-ex",
    doi = "10.1103/PhysRevLett.123.161301",
    journal = "Phys. Rev. Lett.",
    volume = "123",
    number = "16",
    pages = "161301",
    year = "2019"
}

@article{XENON:2018voc,
    author = "Aprile, E. and others",
    collaboration = "XENON",
    title = "{Dark Matter Search Results from a One Ton-Year Exposure of XENON1T}",
    eprint = "1805.12562",
    archivePrefix = "arXiv",
    primaryClass = "astro-ph.CO",
    doi = "10.1103/PhysRevLett.121.111302",
    journal = "Phys. Rev. Lett.",
    volume = "121",
    number = "11",
    pages = "111302",
    year = "2018"
}

@article{SuperCDMS:2015eex,
    author = "Agnese, R. and others",
    collaboration = "SuperCDMS",
    title = "{New Results from the Search for Low-Mass Weakly Interacting Massive Particles with the CDMS Low Ionization Threshold Experiment}",
    eprint = "1509.02448",
    archivePrefix = "arXiv",
    primaryClass = "astro-ph.CO",
    reportNumber = "IPPP-15-56, DCTP-15-112, FERMILAB-PUB-15-394-AE",
    doi = "10.1103/PhysRevLett.116.071301",
    journal = "Phys. Rev. Lett.",
    volume = "116",
    number = "7",
    pages = "071301",
    year = "2016"
}

@article{SuperCDMS:2013eoh,
    author = "Agnese, R. and others",
    collaboration = "SuperCDMS",
    title = "{Search for Low-Mass Weakly Interacting Massive Particles Using Voltage-Assisted Calorimetric Ionization Detection in the SuperCDMS Experiment}",
    eprint = "1309.3259",
    archivePrefix = "arXiv",
    primaryClass = "physics.ins-det",
    reportNumber = "FERMILAB-PUB-13-572-AE",
    doi = "10.1103/PhysRevLett.112.041302",
    journal = "Phys. Rev. Lett.",
    volume = "112",
    number = "4",
    pages = "041302",
    year = "2014"
}

@article{Hertel:2018aal,
    author = "Hertel, S. A. and Biekert, A. and Lin, J. and Velan, V. and McKinsey, D. N.",
    title = "{Direct detection of sub-GeV dark matter using a superfluid $^4$He target}",
    eprint = "1810.06283",
    archivePrefix = "arXiv",
    primaryClass = "physics.ins-det",
    doi = "10.1103/PhysRevD.100.092007",
    journal = "Phys. Rev. D",
    volume = "100",
    number = "9",
    pages = "092007",
    year = "2019"
}

@article{SuperCDMS:2016wui,
    author = "Agnese, R. and others",
    collaboration = "SuperCDMS",
    title = "{Projected Sensitivity of the SuperCDMS SNOLAB experiment}",
    eprint = "1610.00006",
    archivePrefix = "arXiv",
    primaryClass = "physics.ins-det",
    reportNumber = "FERMILAB-PUB-16-467-AE",
    doi = "10.1103/PhysRevD.95.082002",
    journal = "Phys. Rev. D",
    volume = "95",
    number = "8",
    pages = "082002",
    year = "2017"
}

@inproceedings{Alexander:2016aln,
    author = "Alexander, Jim and others",
    title = "{Dark Sectors 2016 Workshop: Community Report}",
    eprint = "1608.08632",
    archivePrefix = "arXiv",
    primaryClass = "hep-ph",
    reportNumber = "FERMILAB-CONF-16-421",
    month = "8",
    year = "2016"
}

@article{Hasegawa:2019jsa,
    author = "Hasegawa, Takuya and Hiroshima, Nagisa and Kohri, Kazunori and Hansen, Rasmus S. L. and Tram, Thomas and Hannestad, Steen",
    title = "{MeV-scale reheating temperature and thermalization of oscillating neutrinos by radiative and hadronic decays of massive particles}",
    eprint = "1908.10189",
    archivePrefix = "arXiv",
    primaryClass = "hep-ph",
    reportNumber = "KEK-TH-2149, KEK-Cosmo-242, RIKEN-iTHEMS-Report-19, IPMU19-0120",
    doi = "10.1088/1475-7516/2019/12/012",
    journal = "JCAP",
    volume = "12",
    pages = "012",
    year = "2019"
}

@inproceedings{Batell:2022dpx,
    author = "Batell, Brian and Blinov, Nikita and Hearty, Christopher and McGehee, Robert",
    title = "{Exploring Dark Sector Portals with High Intensity Experiments}",
    booktitle = "{Snowmass 2021}",
    eprint = "2207.06905",
    archivePrefix = "arXiv",
    primaryClass = "hep-ph",
    month = "7",
    year = "2022"
}

@article{Redmond:2017tja,
    author = "Redmond, Kayla and Erickcek, Adrienne L.",
    title = "{New Constraints on Dark Matter Production during Kination}",
    eprint = "1704.01056",
    archivePrefix = "arXiv",
    primaryClass = "hep-ph",
    doi = "10.1103/PhysRevD.96.043511",
    journal = "Phys. Rev. D",
    volume = "96",
    number = "4",
    pages = "043511",
    year = "2017"
}

@article{DEramo:2017gpl,
    author = "D'Eramo, Francesco and Fernandez, Nicolas and Profumo, Stefano",
    title = "{When the Universe Expands Too Fast: Relentless Dark Matter}",
    eprint = "1703.04793",
    archivePrefix = "arXiv",
    primaryClass = "hep-ph",
    reportNumber = "SCIPP-17-02",
    doi = "10.1088/1475-7516/2017/05/012",
    journal = "JCAP",
    volume = "05",
    pages = "012",
    year = "2017"
}

@article{Visinelli:2017qga,
    author = "Visinelli, Luca",
    title = "{(Non-)thermal production of WIMPs during kination}",
    eprint = "1710.11006",
    archivePrefix = "arXiv",
    primaryClass = "astro-ph.CO",
    reportNumber = "NORDITA-2017-114",
    doi = "10.3390/sym10110546",
    journal = "Symmetry",
    volume = "10",
    number = "11",
    pages = "546",
    year = "2018"
}

@article{Thrane:2013oya,
    author = "Thrane, Eric and Romano, Joseph D.",
    title = "{Sensitivity curves for searches for gravitational-wave backgrounds}",
    eprint = "1310.5300",
    archivePrefix = "arXiv",
    primaryClass = "astro-ph.IM",
    doi = "10.1103/PhysRevD.88.124032",
    journal = "Phys. Rev. D",
    volume = "88",
    number = "12",
    pages = "124032",
    year = "2013"
}

@article{Caprini:2015zlo,
    author = "Caprini, Chiara and others",
    title = "{Science with the space-based interferometer eLISA. II: Gravitational waves from cosmological phase transitions}",
    eprint = "1512.06239",
    archivePrefix = "arXiv",
    primaryClass = "astro-ph.CO",
    reportNumber = "DESY-15-246",
    doi = "10.1088/1475-7516/2016/04/001",
    journal = "JCAP",
    volume = "04",
    pages = "001",
    year = "2016"
}

@article{Steigman:2012nb,
    author = "Steigman, Gary and Dasgupta, Basudeb and Beacom, John F.",
    title = "{Precise Relic WIMP Abundance and its Impact on Searches for Dark Matter Annihilation}",
    eprint = "1204.3622",
    archivePrefix = "arXiv",
    primaryClass = "hep-ph",
    doi = "10.1103/PhysRevD.86.023506",
    journal = "Phys. Rev. D",
    volume = "86",
    pages = "023506",
    year = "2012"
}

@article{Arcadi:2017kky,
    author = "Arcadi, Giorgio and Dutra, Ma{\'\i}ra and Ghosh, Pradipta and Lindner, Manfred and Mambrini, Yann and Pierre, Mathias and Profumo, Stefano and Queiroz, Farinaldo S.",
    title = "{The waning of the WIMP? A review of models, searches, and constraints}",
    eprint = "1703.07364",
    archivePrefix = "arXiv",
    primaryClass = "hep-ph",
    doi = "10.1140/epjc/s10052-018-5662-y",
    journal = "Eur. Phys. J. C",
    volume = "78",
    number = "3",
    pages = "203",
    year = "2018"
}

@article{Roszkowski:2017nbc,
    author = "Roszkowski, Leszek and Sessolo, Enrico Maria and Trojanowski, Sebastian",
    title = "{WIMP dark matter candidates and searches{\textemdash}current status and future prospects}",
    eprint = "1707.06277",
    archivePrefix = "arXiv",
    primaryClass = "hep-ph",
    reportNumber = "UCI-HEP-TR-2017-09, DO-TH-17-15, UCI-HEP-TR-2017-09-",
    doi = "10.1088/1361-6633/aab913",
    journal = "Rept. Prog. Phys.",
    volume = "81",
    number = "6",
    pages = "066201",
    year = "2018"
}

@article{Arcadi:2024ukq,
    author = "Arcadi, Giorgio and Cabo-Almeida, David and Dutra, Ma{\'\i}ra and Ghosh, Pradipta and Lindner, Manfred and Mambrini, Yann and Neto, Jacinto P. and Pierre, Mathias and Profumo, Stefano and Queiroz, Farinaldo S.",
    title = "{The Waning of the WIMP: Endgame?}",
    eprint = "2403.15860",
    archivePrefix = "arXiv",
    primaryClass = "hep-ph",
    doi = "10.1140/epjc/s10052-024-13672-y",
    journal = "Eur. Phys. J. C",
    volume = "85",
    number = "2",
    pages = "152",
    year = "2025"
}

@article{XENON:2024wpa,
    author = "Aprile, E. and others",
    collaboration = "XENON",
    title = "{The XENONnT dark matter experiment}",
    eprint = "2402.10446",
    archivePrefix = "arXiv",
    primaryClass = "physics.ins-det",
    doi = "10.1140/epjc/s10052-024-12982-5",
    journal = "Eur. Phys. J. C",
    volume = "84",
    number = "8",
    pages = "784",
    year = "2024"
}

@article{LZ:2024zvo,
    author = "Aalbers, J. and others",
    collaboration = "LZ",
    title = "{Dark Matter Search Results from 4.2{\,}{\,}Tonne-Years of Exposure of the LUX-ZEPLIN (LZ) Experiment}",
    eprint = "2410.17036",
    archivePrefix = "arXiv",
    primaryClass = "hep-ex",
    reportNumber = "FERMILAB-PUB-24-0796-V",
    doi = "10.1103/4dyc-z8zf",
    journal = "Phys. Rev. Lett.",
    volume = "135",
    number = "1",
    pages = "011802",
    year = "2025"
}

@article{PandaX:2025rrz,
    author = "Zhang, Minzhen and others",
    collaboration = "PandaX",
    title = "{Search for Light Dark Matter with 259 Days of Data in PandaX-4T}",
    eprint = "2507.11930",
    archivePrefix = "arXiv",
    primaryClass = "hep-ex",
    doi = "10.1103/rtnh-jn8s",
    journal = "Phys. Rev. Lett.",
    volume = "135",
    number = "21",
    pages = "211001",
    year = "2025",
    note = "[Erratum: Phys.Rev.Lett. 136, 069901 (2026)]"
}
\end{document}